\documentclass[twocolumn]{aastex63}
\usepackage{amsmath}
\usepackage[figuresright]{rotating}
\usepackage{color}

\newcommand{\msol}{${\rm M}_{\odot}$}
\newcommand{\feh}{${\rm [Fe/H]}$}
\newcommand{\alphafe}{$[\alpha/{\rm Fe}]$}
\newcommand{\fehalpha}{$([{\rm Fe/H}], [\alpha/{\rm Fe}])$}

\def\be{\begin{eqnarray}}   \def\ee{\end{eqnarray}}

\begin{document}

\title{Chemical Evolution with Radial Mixing Redux: A Detailed Model for Formation and Evolution of the Milky Way}


\author[0000-0001-7083-2417]{Boquan Chen}
\affiliation{Sydney Institute for Astronomy, School of Physics, The University of Sydney, NSW 2006, Australia}
\affiliation{ARC Centre of Excellence for All Sky Astrophysics in Three Dimensions (ASTRO-3D)}
\author[0000-0001-7294-9766]{Michael R. Hayden}
\affiliation{Sydney Institute for Astronomy, School of Physics, The University of Sydney, NSW 2006, Australia}
\affiliation{ARC Centre of Excellence for All Sky Astrophysics in Three Dimensions (ASTRO-3D)}
\author[0000-0002-0920-809X]{Sanjib Sharma}
\affiliation{Sydney Institute for Astronomy, School of Physics, The University of Sydney, NSW 2006, Australia}
\affiliation{ARC Centre of Excellence for All Sky Astrophysics in Three Dimensions (ASTRO-3D)}
\author[0000-0001-7516-4016]{Joss Bland-Hawthorn}
\affiliation{Sydney Institute for Astronomy, School of Physics, The University of Sydney, NSW 2006, Australia}
\affiliation{ARC Centre of Excellence for All Sky Astrophysics in Three Dimensions (ASTRO-3D)}
\author[0000-0002-4343-0487]{Chiaki Kobayashi}
\affiliation{Centre for Astrophysics Research, Department of Physics, Astronomy and Mathematics, University of Hertfordshire, Hatfield, AL10 9AB UK}
\affiliation{ARC Centre of Excellence for All Sky Astrophysics in Three Dimensions (ASTRO-3D)}
\author[0000-0002-3625-6951]{Amanda I. Karakas}
\affiliation{School of Physics \& Astronomy, Monash University, Clayton Vic 3800, Australia}
\affiliation{ARC Centre of Excellence for All Sky Astrophysics in Three Dimensions (ASTRO-3D)}

\begin{abstract}
{We present a multi-zone galactic chemical evolution (GCE) model for the Milky Way that takes the most recently updated yields of major nucleosynthesis channels into account. It incorporates physical processes commonly found in previous GCE models like supernova and star formation feedback, the radial flow of gas in the disk, and the infall of fresh gas, along with stellar scattering processes like radial migration. We individually analyse the effect of different physical processes present in our model on the observed properties of the Galaxy. The radial flow of gas in the disk plays an important role in establishing the radial gradient for \feh{} in the low-\alphafe{} sequence. Our model with one episode of smooth gas infall and constant star formation efficiency is capable of reproducing the observed \fehalpha{} distribution of stars at different ($R$, $|z|$) positions in the Milky Way. Our results point to the rapid evolution of \alphafe{} after the onset of Type Ia supernovae and a high star formation rate during the formation of the high-\alphafe{} sequence as the origin of dual peaks in \alphafe{}. A secondary infall is unnecessary to reproduce the \alphafe{}-gap and chemical spread in the disk in our model. We additionally compare the median age for various mono-abundance populations and the age-metallicity relation at different ($R$, $|z|$) positions from our fiducial model to observations. We discuss our results in relation to other related work in detail. }

\end{abstract}

\section{Introduction} \label{sec:intro}
The formation and evolution of galaxies are fundamental questions confronting astrophysics today. In particular, the detailed chemical enrichment history of galaxies involves a complex narrative that is poorly understood. The Milky Way provides direct observational evidence of how chemical abundances evolve with time from individual stars and stellar populations. Stars contain the chemical imprint of the gas from which they formed, acting as a powerful tracer of the Milky Way's chemical evolution in space and time \citep{2002ARA&A..40..487F}, provided that we can accurately estimate stellar ages. Galactic chemical evolution (GCE) models make assumptions about the general physical processes that govern the evolution of galaxies and approximate them with empirical laws. These models can be used to simulate the chemical enrichment history of the Galaxy in a short amount of time if properly constrained by all of the available stellar abundance data.


The simplest GCE model assumes that a galaxy can be treated as one zone \citep{1962AJ.....67..486V, 1963ApJ...137..758S}. The zone is initially filled with gas that is pristine (containing only H, He, and a trace amount of Li from Big Bang nucleosynthesis). As stars form out of available gas, heavier elements produced by stellar nucleosynthesis enrich the existing gas and a chemical evolutionary track that traces the abundance patterns over time is generated. Infalling gas from an intergalactic medium (IGM) whose composition does not necessarily match that of the existing gas can be added to the model over time to sustain star formation and moderate the rate of chemical evolution. 

The one-zone model can be extended into a multi-zone model by assuming that a galaxy is composed of concentric rings \citep{1997ApJ...477..765C, 2001ApJ...554.1044C}. As a number of factors vary in each zone, such as gas density, inflow from the IGM, and supernova-heated outflow, the model draws differential evolutionary tracks from the innermost to the outermost region. Multi-zone models are much more complex because they require additional assumptions that control the distribution of materials across different radii over time and allow mechanisms that exchange gas and stars among the zones. Observations of other galaxies suggest that disk galaxies experience inside-out growth \citep{2014ApJ...788...28V, 2017MNRAS.470..651R}. The Milky Way also showed signs of inside-out growth \citet{2012ApJ...753..148B}. The growth in the half-mass radius of the Milky Way is quantified to be 43\% over the last 7 Gyr \citep{2019ApJ...884...99F}. Both inside-out growth (e.g.,\citealt{1989MNRAS.239..885M}) and radial flow (radial movement of gas in the disk, e.g., \citealt{2009MNRAS.396..203S}), can prove crucial to generating models that match observations as chemically diverse stars in different locations of the Galaxy do not have to be born in-situ and be explained by various physical conditions elsewhere.

There are many GCE models in the literature. One famous model is the two-infall model by \cite{1997ApJ...477..765C} which assumes two distinct infall episodes. The first episode happens at the beginning of the model, which helps form the halo and the \alphafe{}-enhanced thick disk stars. {After the gas density falls below the star formation threshold (7 \msol{} per $\mathrm{pc^2}$),} the second episode brings in a large amount of fresh gas to sustain star formation for a long period of time. The second infall helps form a large amount of metal-rich stars to match the stellar distribution of metallicity in the Milky Way, compared to a closed-box model. Since the two peaks in the stellar distribution of \alphafe{} have been identified, the two-infall model has been modified to explain this feature of our Galaxy. An updated two-infall model by \citet{2019A&A...623A..60S} placed a delay of about four Gyr between the two infalls and its chemical track is characterised by a loop (drop in \feh{} and increase in \alphafe{}) in the \fehalpha{} plane. The thin disk gets chemically enriched over time by the second infall to reproduce the large spread in \feh{} in the low-\alphafe{} sequence.

Depending on which part of the Milky Way we study, there are two terms to describe the two density peaks in \alphafe{}. The \alphafe{}-bimodality refers to the two \alphafe{}-peaks in the bulge. Most stars in the bulge are born in situ. The \alphafe{}-dichotomy refers to the two \alphafe{}-peaks seen away from the bulge, such as the solar neighbourhood. The high-\alphafe{} stars in these regions do not necessarily form in situ and radial migration could play a significant role. \cite{2018A&A...618A..78H} found that a decrease in the SFR (quenching) driven by a rapid drop in the SFE and gas accretion rate during the intermediate-\alphafe{} regime is required for the \alphafe{}-bimodality. Similar to the two-infall models, they proposed a scenario where the low-\alphafe{} sequence formed from chemically distinct gas with some delay. Then, \cite{2019A&A...625A.105H} argued that the chemical gradient in the disk is caused by the dilution of the pre-enriched gas from the formation of the high-\alphafe{} sequence by the metal-poor fresh gas in the Galaxy. Since the high-\alphafe{} stars primarily formed in the inner disk, the polluted ISM becomes more metal-poor and high-\alphafe{} as we move away from the Galactic centre. Inspired by their work, \cite{2020MNRAS.497.2371L} found that both the quenching episode and the second infall episode are necessary for the \alphafe{}-bimodality with their GCE model and estimated the delay between the two infall episodes to about six Gyr. Due to the long delay, a significant amount of metal-rich stars formed in the bulge before the second infall in their model.

One commonly cited model that takes into account the effect of radial migration is \cite{2009MNRAS.396..203S}. They showed that the extended \feh{} distribution of the thin-disk (low-\alphafe{}) stars in the \fehalpha{} plane is due to radial migration and one smooth episode of gas infall is sufficient to generate the \alphafe{}-dichotomy in the disk. \cite{2013A&A...558A...9M} harnessed the complex dynamical processes from cosmological simulations and combined them with a pure GCE model, concluding that mergers play an important role in the formation of the thick disk. Similarly, \cite{2015A&A...580A.126K} was able to reproduce most properties of the Milky Way disk by modelling the growth of the disk and the effect of radial migration based on simulations. It is common for some GCE models today to draw some aspects of their models from simulations. The details in simulations can even become the basis of a 2D GCE model that allows us to study azimuthal variations \citep{2019A&A...628A..38S}. However, directly importing particle properties from cosmological simulations often results in a very noisy picture of the replicated galaxies and can obscure the conclusions we can draw from the chemical properties.

One key ingredient of GCE models is chemical yields from stellar nucleosynthesis, i.e., the amount of newly synthesized elements released into the ISM upon the death of a star. The chemical yields are largely dependent on the mass and metallicity of the progenitors. Stars with an initial mass of between ~$\sim 0.5 - 7 \rm \rm M_\odot$ become asymptotic giant branch (AGB) stars and experience mass loss towards the end of their lifetime through stellar winds, leaving behind C+O white dwarfs (WDs). The planetary nebulae AGB stars release primarily carbon, nitrogen, oxygen and small amounts of $s$-process elements. $s$-process elements include about half the elements heavier than iron. Massive stars of $10-40 \rm M_\odot$ form core-collapse supernovae (CCSN). The explosive energy produces $\alpha$ elements, i.e. C, O, Ne, Mg, Si, S, Ar and Ca through alpha capture and $r$-process elements through rapid neutron capture which is responsible for the other half of elements heavier than iron. The third major channel of metal production is Type Ia supernova (SNe Ia). The SNe Ia yields for single degenerate progenitors where a white dwarf progenitor explodes when its mass reaches Chandrasekhar limit ($\sim 1.4 \rm M_\odot$) are commonly adopted \citep{1999ApJS..125..439I}. SNe Ia primarily produces iron-peak elements. \cite{2013ARA&A..51..457N} discusses these three production channels and others in greater detail. Tremendous improvements have been made to nucleosynthesis yields in recent years. In this work, we utilize the yields from \cite{2020ApJ...900..179K} and \cite{2020ApJ...895..138K} which contain up to 83 elements, more than any current spectroscopic surveys can measure.

In recent years, an unprecedented amount of details of the multi-elemental abundance distribution in the Milky Way are being revealed by large-scale spectroscopic surveys, e.g. GALactic Archaeology with HERMES (GALAH) \citep{2021MNRAS.tmp.1259B}, Apache Point Observatory Galactic Evolution Experiment (APOGEE) \citep{2020ApJS..249....3A}, and Large Sky Area Multi-Object Fibre Spectroscopic Telescope (LAMOST) \citep{2012RAA....12.1243L}. These surveys aim to observe large samples of stars ($\gtrsim1$ million), with multiple chemical abundances, and with good spatial coverage of the Galaxy. The chemical information of these stars is further enhanced by the kinematic properties provided by the Gaia mission \citep{2021A&A...649A...1G}. The chemical and kinematic properties both at the solar radius and beyond give us an unprecedented amount of information to constrain the parameters of GCE models. \citet{2015ApJ...808..132H} using APOGEE data studied the \fehalpha{} distribution of stars at different Galactocentric $(R,z)$ locations of the Galaxy. They found a clear presence of two distinct sequences one with high \alphafe{} and another with low \alphafe{}. In general, the high \alphafe{} sequence was dominant at low $R$ and high $|z|$. However, at large $R$ and large $|z|$ the low \alphafe{} sequence was dominant. Explaining these trends with GCE models requires a match simultaneously in different ($R$, $|z|$) positions and proves rather challenging.

Recently, simulations of Milky-Way-type galaxies are successful at replicating the dual \alphafe{}-peaks observed in the Galaxy. \cite{2019MNRAS.484.3476C} showed that a high star formation rate caused by the fragmentation of an early gas-rich disk can naturally give rise to a thin and thick disk with different \alphafe{} in non-cosmological simulations. \citet{2021MNRAS.501.5176K} also emphasized the importance of a high star formation rate for the formation of the high-\alphafe{} sequence but added that the feedback from this early star formation episode leading to a rapid shutdown of star formation is crucial to a dichotomy of \alphafe{}. \citet{2020MNRAS.491.5435B} showed that a gas-rich merger that lowers metallicity and leaves \alphafe{} untouched can induce an \alphafe{}-dichotomy in NIHAO-UHD simulations, similar to what a two-infall model proposes. \cite{2021MNRAS.503.5826A} and \cite{2021MNRAS.503.5868R} used an enriched source of gas to feed the formation of the high-\alphafe{} sequence in the inner parts of a galaxy and a more metal-poor source for the outer parts to create a bimodal-\alphafe{} distribution. There is much evidence from simulations that accretion with a customized chemical composition can induce two \alphafe{}-sequences. However, it does not eliminate the possibility that the two \alphafe{}-sequences can arise without external factors. Although simulations could generate two density peaks in the distribution of \alphafe{}, many finer details in the \fehalpha{} plane, such as the \alphafe{}-gap and the \feh{} gradient, seldom match what we observe in the Milky Way. It is very computationally expensive to implement detailed modelling of nucleosynthesis in simulations.

\citet{2021MNRAS.507.5882S} presented a pure analytic chemodynamical model of the Milky Way that was able to replicate the observed abundance trends of \citet{2015ApJ...808..132H} with simple analytic approximations of just a few physical processes. They showed that velocity dispersion relations and radial migration play an important role in shaping the abundance maps and their variation across $(R,|z|)$. The most important factor for the origin of the two sequences was related to the sharp fall of \alphafe{} with time which suppressed star formation in the intermediate-\alphafe{} region. In order to replicate the stellar density distribution in the \fehalpha{} plane, both \alphafe{} and \feh{} have to evolve rapidly with time early on but reach equilibrium values not long after the thin disk formed. The radial and vertical motion of stars that also evolve with time are sufficient to explain the radial gradient in both \feh{} and \alphafe{} with time. The question we would like to answer is whether a chemical evolution model derived from the first principles (taking into account star formation and nucleosynthesis yields) can lead to the same evolution of \alphafe{} and \feh{} with time as seen by the analytic model presented in \citet{2021MNRAS.507.5882S}. Additionally, we want to identify the processes that can speed up or slow down the fall of \alphafe{} with time and those that determine the equilibrium value of abundances in the disc as well as the radial gradient of abundances. 

{The work by \citet{2009MNRAS.396..203S} is groundbreaking because it does not assume that the stars with a wide range of metallicity values are born in the solar neighbourhood. Instead, stars with different metallicities are born at corresponding radii according to the metallicity gradient in the disk and migrate throughout the disk to generate the spread in metallicity. The advent of large-scale spectroscopic surveys, such as APOGEE and GALAH, combined with the precise proper motion and parallax measurements from \textit{Gaia}, provide an unprecedented picture of our Galaxy. We now have a more complete picture of the chemical distribution beyond the solar neighbourhood in the disk for many elemental abundances. We present our model and results as a much-needed update to the original work by \citet{2009MNRAS.396..203S}.} The paper is organized as follows. In Section \ref{sec:ingredients} we describe the basic ingredients of our model. Other than our fiducial model, we also create alternate models with different parameter values and mechanisms to help readers form some intuition on their effects on the final observable. The resulting chemical tracks and stellar density distributions in \feh{} and \alphafe{} of these models are shown in Section \ref{sec:results}. In Section \ref{sec:discussion}, we discuss the implications of our results in comparison to other relevant work. Finally, we summarize and present our main conclusions in Section \ref{sec:conclusion}.



\section{Basic Ingredients} \label{sec:ingredients}

\subsection{Setting up basic framework} \label{subsec:setting up}

Our model is built upon \cite{2009MNRAS.396..203S} and uses the same setup. The disk in our model consists of rings (radial zones or shells) that are $\Delta R = 0.25$ kpc wide with central radii ranging from 0.125 kpc to 19.875 kpc. Each zone contains a cold gas, warm gas, and stellar component. All three record the mass of each element present in the nucleosynthesis yield tables. Stars form only out of cold gas, locking in the gas of the same composition. Warm gas captures the majority of enriched gas from nucleosynthesis and releases it slowly with a timescale $t_{\rm cool}=1.2$ Gyr to the cold gas for future star formation. The stellar component in each zone keeps track of the current guiding radii of stars at each time step and distributes yields accordingly. The model lasts for a total of 12 Gyr. Each time step advances the model forward by $\Delta t =$ 30 Myr, which roughly corresponds to the maximum lifetime of the least massive CCSN progenitors and the free-fall time of molecular gas \citep{2022AJ....164...43S}.

We initialize the disk with $M_{0, \rm cold} = 7.0 \times 10^9 {\rm M}_\odot$ of cold gas and $M_{0, \rm warm} = 1.0 \times 10^9 {\rm M}_\odot$ of warm gas, both with a pristine composition containing only H, He, and a trace amount of Li. The initial amount of cold gas determines the amount of star formation in the first step and in turn the starting \feh{} in the model. However, it has a limited effect on our model because a large amount of infalling gas also enters our model as cold gas from the first step. We chose the parameter values so that our fiducial model could have a relatively high starting \feh{} to reach \feh{}=-1 in about one Gyr and still replicate the stellar density distribution in the \fehalpha{}-plane. The cold and warm gas initially present as well as the cold gas at the beginning of every time step in each zone is restricted to a pre-determined radial mass distribution which is the result of an assumed exponential density profile with a scalelength of $R_d=$ 3.5 kpc (Equation \ref{eq:density}). The exact mass fraction in each zone can be calculated with the integral in Equation \ref{eq:mass_frac} where $f_{n, \rm cold}$ is the mass fraction in the n-th zone and $R_n$ is its central radius. 
\begin{equation} \label{eq:density}
    \Sigma_{\mathrm{cold}, R} \sim e^{-R/ R_d}
\end{equation}

\begin{equation} \label{eq:mass_frac}
    f_{n, \rm cold} = \int_{R_n - \Delta R}^{R_n + \Delta R} e^{-R/ \rm R_d} dR
\end{equation}

Theoretically, the scalelength of the gas disk in our model can change over time to accommodate an inside-out growth scenario. However, we found that changing the scalelength even a little introduced a large amount of instability to the chemical evolutionary tracks even after the abundances approached equilibrium values. We decided to keep our scalelength constant as \cite{2009MNRAS.396..203S} in favour of exploring the effect of other parameters for this work. During each of the four time steps immediately after the initial two steps (60 Myr), an additional $\Delta M_{\rm warm} = 2.5 \times 10^8 \rm M_\odot$ with the mass-weighted average composition of the warm ISM in the disk is added onto the warm disk. This early infall is distributed based on the same density profile in Eq \ref{eq:density} and helps form the observed amount of metal-poor halo stars. We found that this brief infall episode had a negligible impact on our final results, but we decided to include it as part of the original model in case more observational evidence on the in situ halo stars were discovered in the future.

\subsection{Star formation} \label{subsec:imf&sfr}

We adopt Kennicutt-Schimidt (KS) law \citep{1998ARA&A..36..189K} for determining star formation mass and calculate the SFR as in Equation \ref{eq:sfr}. The surface density of cold gas $\Sigma_{\mathrm{cold}}$ is measured in $M_{\odot} / {\rm pc}^2$ while time $t$ is in yr.
\begin{equation} \label{eq:sfr}
    \frac{d\Sigma_*}{dt} = C \Sigma_{\mathrm{cold}}^{1.4}
\end{equation}
where $C$ is a constant equivalent to the star formation efficiency (SFE) in the model. The combination of a constant scalelength ($R_d=$ 3.5 kpc) for the cold gas and KS law with a power of 1.4 results in a stellar disc with scalelength  $R_d/1.4 = 2.5$ kpc. In our fiducial model, we adopt $C=2.8 \times 10^{-10}$ as our SFE, close to $2.5 \times 10^{-10}$ estimated from the Hubble sequence by \citet{1998ARA&A..36..189K}. This number is about 2.5 times higher than that adopted by \cite{2009MNRAS.396..203S} ($1.2 \times 10^{-10}$) because we found that a high star formation rate (SFR) is essential for building a high stellar count for the high-\alphafe{}-sequence and for \alphafe{} to drop rapidly during the intermediate-\alphafe{} phase. Our adopted value is lower than of \cite{2008AJ....136.2846B} and \cite{2008AJ....136.2782L} ($5 \times 10^{-10}$) because their value is for the molecular gas, which is a fraction of the cold ISM. Besides the fiducial model, we created two additional models named \textit{SF-} (a lower SFE) and \textit{SF+} (a higher SFE) with $C=1.2 \times 10^{-10}$ and $C=4.4 \times 10^{-10}$ respectively to explore the effect of the SFE. 

We adopt \cite{Kroupa_2001} initial mass function (IMF), given by
\begin{align} \label{eq:imf}
    \xi(m) \propto
    \begin{cases}
            m^{-1.3} & \mathrm{0.1 \leq m < 0.5 \rm M_\odot} \\
            m^{-2.3} & \mathrm{m \geq 0.5 \rm M_\odot}.
    \end{cases}
\end{align}
which is commonly used in chemical evolution models. We restrict the mass range to $(0.1, 50) \rm M_\odot$. The width of the mass bins is 0.1 $\rm {M_{\odot}}$ for bins up to $m_{{\mathrm{cutoff}}} = 9 \ {M_{\odot}}$ and 1 ${M_{\odot}}$ for bins with $m \geq m_{{\mathrm{cutoff}}}$. Changing the bounds of stellar bins did not have any significant effect on the final chemical tracks. The IMF is normalized such that $\int m\xi(m) dm = 1 \ {M_{\odot}}$. The average mass of each bin is calculated as $\int_{m_0}^{m_1} m\xi(m) dm / \int_{m_0}^{m_1} \xi(m) dm$, where $m_0$ and $m_1$ are the lower and the upper bounds of the said bin. 

\subsection{Stellar lifetime} \label{subsec:age}

\begin{figure}
  \includegraphics[width=\linewidth]{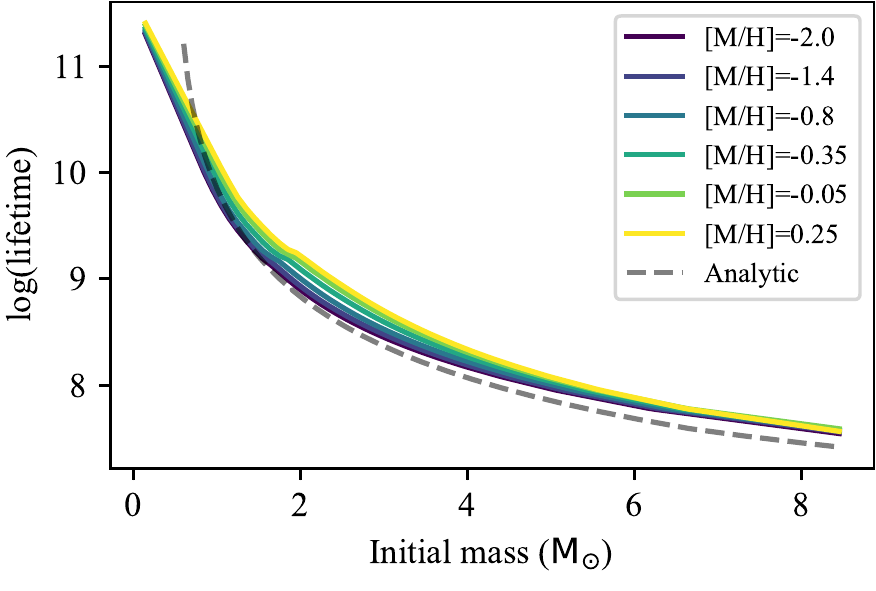}
  \vspace*{-6mm}
  \caption{Stellar lifetimes as a function of initial stellar mass for different metallicities  based on PARSEC-1.2S isochrones by metallicity \citep{2012MNRAS.427..127B}. The dashed line shows the analytical formula adopted by \cite{1993ApJ...416...26P} in their study. }
  \label{fig:isochrone_age}
\end{figure}

The stellar lifetimes for a given progenitor mass and metallicity are extracted from PARSEC-1.2S isochrones \citep{2012MNRAS.427..127B}. We downloaded isochrones with ages identical to the time steps in our model from 0 to 12 Gyr at [M/H] from -2 to 0.5 dex in increments of 0.1 dex below [M/H]=-0.1 and 0.05 dex otherwise from \url{http://stev.oapd.inaf.it/cgi-bin/cmd}.
For each isochrone, we record its age, metallicity and maximum mass excluding white dwarfs to create a new table. Then we interpolate over this table to determine the stellar lifetime for any given mass and metallicity.
Figure \ref{fig:isochrone_age} shows the stellar lifetimes for various progenitor masses at different metallicity values. We added the analytic stellar age from \cite{1993ApJ...416...26P} as a dashed line for comparison. As stellar age does not vary greatly with metallicity, we choose not to interpolate over the metallicity grid to save computational costs. Our model uses the stellar lifetime table with the closest metallicity instead. 

\begin{table*}
\caption{Parameters of the fiducial model.}              
\label{table:parameters}      
\centering                                      
\begin{tabular}{l l l l}          
\hline\hline                        
Parameter & Meaning & Value \\    
\hline                                   
    $R_d$ & Scalelength of cold gas & 3.5 kpc  \\
    $C$ & Star formation efficiency constant & $2.8 \times 10^{-10}$ \\
    $N$ & Power in star formation law & 1.4 \\
    $M_{0, \rm cold}$ & Initial cold gas mass  & $7 \times 10^9 \rm M_\odot$ \\    
    $M_{0, \rm warm}$ & Initial warm gas mass  & $1 \times 10^9 \rm M_\odot$    \\
    $\Delta M_{\rm warm}$ & Early infall mass per step onto warm gas & $2.5 \times 10^8 \rm M_\odot$ \\
    $M_1$ & Short timescale infall mass & $4 \times 10^{10} \rm M_\odot$       \\
    $M_2$ & Long timescale infall mass & $7 \times 10^{10} \rm M_\odot$     \\
    $b_1$ & Timescale for $M_1$ & 150 Myr    \\
    $b_2$ & Timescale for $M_2$ & 1.4 Gyr    \\
    $t_{\rm min, SNeIa}$ & Minimum time delay before first SNe Ia & 150 Myr    \\
    $t_{\rm scale, SNeIa}$ & Timescale for decay of SNe Ia & 1.5 Gyr \\
    $t_{\rm cool}$ & Cooling timescale of warm gas & 1.2 Gyr   \\
    $f_{\rm direct}$ & Fraction of supernovae ejecta directly into cold gas & 0.01     \\
    $f_{\rm eject}$ & Fraction of supernovae ejecta lost & 0.2  \\
    $\eta_{\rm SF}$ & Multiplier for gas heated by star formation & 1.0 \\
    $\eta_{\rm SN}$ & Multiplier for gas heated by supernovae & 3.0 \\
    $Z_{\rm M_\odot}$ & Metallicity of the Sun & 0.0156 \\
    $Z_{\rm IGM}$ & Metallicity of the IGM & $10^{-0.8} Z_{\rm M_\odot}$ \\
    $f_{\rm IGM}$ & The fraction of gas directly deposited from IGM & 0.3     \\
    \%SNeIa & Fraction of white dwarfs from stars within (3.2, 8.5)$\rm M_\odot$ that turn into SNe Ia & 0.65-0.05 \\
    \%HNe & Fraction of CCSN that explode as hypernovae & 0.5\\
    \%MRSN & Fraction of CCSN that explode as magneto-rotational supernovae & 0.0025 \\
    $v_0$ & Rotational velocity of the Milky Way from \cite{2015MNRAS.449.3479S} & 220 km/s\\
    $\sigma_{L_0}$ & Churning amplitude from \cite{2015MNRAS.449.3479S} & 1150 kpc km/s  \\
    $t_{max}$ & Maximum age of the Milky Way & 12 Gyr\\
\hline    
\end{tabular}
\end{table*}

\subsection{Stellar yields} \label{subsec:yield}
We adopt the state-of-the-art yields of major nucleosynthesis production sites from \cite{2020ApJ...900..179K}, including AGB, CCSN, and SNe Ia. The CCSN includes hypernovae (HNe), Type II supernovae (SNe II) and magneto-rotational supernovae (MRSNe). The yields for MRSNe were adopted from 
\cite{2015ApJ...810..109N} (publicly available at \url{https://www.astro.keele.ac.uk/~nobuya/mrsn/}) and the yields for the rest were adopted from \cite{2020ApJ...900..179K} and \cite{2020ApJ...895..138K} (provided to us on request). We linearly interpolate the yields first on our mass grids specified in Section \ref{subsec:imf&sfr} and then on a grid of 1,001 metallicity values evenly spaced on a logarithmic scale from $Z=10^{-6}$ to $Z=0.05$. 

The yields from AGB stars are applicable to stars (stellar bins in our model) with an initial mass of 0.9-9 $\rm M_\odot$, with the exact limits depending upon the metallicity. The maximum AGB progenitor mass increases from 7.5 $\rm M_\odot$ at $Z=0$ to 9 $\rm M_\odot$ at $Z = 0.05$. The SNe II yields are applicable to stars with an initial mass of 8-40 $\rm M_\odot$. The minimum SNe II progenitor mass increases from 8 $\rm M_\odot$ at $Z = 0$ to 10 $\rm M_\odot$ at $Z=0.05$. The HNe yields are only applicable to stars with an initial mass of 20-40 $\rm M_\odot$. Consistent with \cite{2020ApJ...900..179K}, half of all SNe II with an initial mass over 20 $\rm M_\odot$ are substituted with HNe, 
(denoted by parameter \%HNe).
\cite{2015ApJ...810..109N} suggest that it is sufficient for 0.1\% of all CCSN to be MRSNe in order to create the amount of r-process material observed in stars today. We substitute 0.2\% of SNe II with MRSNe in order for [Eu/Fe] to reach solar value at the solar radial zone. Thus, MRSNe share the same progenitor mass range as SNe II. 

The AGB and MRSNe yields list 83 elements in \cite{2020ApJ...900..179K}, while the SNe II, HNe, and SNe Ia yield only list elements up to Ge. 
Each row in the tables lists the mass remnant and yields for individual elements for a certain metallicity and progenitor mass. The total amount of mass returned to the system for each AGB is the difference between its progenitor mass and remnant mass. The AGB yield table lists the mass difference of individual elements before and after stellar evolution and thus the sum of each row is zero. As lighter elements are fused into heavier elements, the yields for light elements, especially H, can become negative. Hence, we need to add the mass of each element trapped inside stellar bins at the time of their formation to retrieve the absolute yields or the actual mass of each element to be returned to the system. 

The entries of each row in SNe II and HNe yield tables are never negative. When we add up the mass remnant and the yields for individual elements for each row, the sum is always slightly less than the progenitor mass. We use gas with the composition at stellar birth as unprocessed gas to make up for this mass difference. The MRSNe yield table lists the relative fraction of each element expected in the final processed metal. We multiply these fractions by the difference between progenitor and remnant mass from each row to derive the absolute mass of each element. The SNe Ia tables list the amount of each element expected to be released per event which adds up to the Chandrasekhar mass limit, $M_{\rm Ch} \sim 1.4 {\rm M}_\odot$, as we only consider single degenerate progenitor scenario.

\begin{table*}
\caption{Changes made to alternate models}              
\label{table:alternate}      
\centering                                      
\begin{tabular}{l l l l}          
\hline\hline                        
Acronym & Meaning & Deviation from the standard model \\    
\hline                                   
CSF & Constant star formation & The total mass of cold ISM is conserved at $8.3 \times 10^9$ after infall \\
C09 & Churning 2009 & Uses the churning mechanism from \cite{2009MNRAS.396..203S} \\
DTD+ & Longer delay timescale for SNe Ia & Delay timescale is increased to 2 Gyr \\
DTD- & Shorter delay timescale for SNe Ia & Delay timescale is shortened to 1 Gyr\\
E+ & Higher ejection ratio & $f_{\rm eject} = 0.4$ \\
E- & Lower ejection ratio & $f_{\rm eject} = 0.0$ \\
MRI & Minimum radial inflow & $f_{\rm IGM}=1.0$ and inflow maintains the scalelength\\
SF+ & Higher star formation efficiency & Star formation constant $C$ is increased to $4.4 \times 10^{-10}$ \\
SF- & Lower star formation efficiency & Star formation constant $C$ is decreased to $1.2 \times 10^{-10}$ \\
SNIa+ & Higher fraction of SNe Ia & \%SNIa=6.5\% at every time step \\
SNIa- & Lower fraction of SNe Ia & \%SNIa=5\% at every time step \\
\hline    
\end{tabular}
\end{table*}

\subsection{SNe Ia fraction and delay-time-distribution} \label{subsec:dtd}
All nucleosynthesis yields from a star are released immediately after the death of a star into the gas components, except for those from SNe Ia. Although the exact nature of SNe Ia explosions requires further study, two SNe Ia progenitor scenarios, the double degenerate \citep{1984ApJS...54..335I, 1984ApJ...277..355W} where two carbon-oxygen (CO) white dwarfs (WD) with a combined mass exceeding $M_{Ch}$ leading to explosive carbon burning and single degenerate \citep{1973ApJ...186.1007W} where a single CO WD accretes hydrogen from a companion until it reaches $M_{Ch}$ and explodes, are the most widely accepted. 

In the fiducial model, we adopt the exponential delay-time-distribution from \cite{2009MNRAS.396..203S} which favours the single degenerate scenario. The number of SNe Ia explosions is determined by $M_{\rm WD}/M_{\rm Ch}$, where $M_{\rm WD}$ is the total mass in white dwarfs (from the mass remnant column in SNe Ia yield tables) in a given radial zone at the current time step. As soon as AGBs in the specified mass range evolve and generate WDs, we calculate the mass of each element from the SNe Ia yields and store them inside a reservoir for a minimum of $t_{\rm min, SNeIa} = 150$ Myr, after which period we release them to the warm ISM where the stars happened to reside in the model exponentially with a timescale of $t_{\rm scale, SNeIa}=1.5$ Gyr. A fraction of the total mass of the remaining WDs equal to $\Delta t/t_{\rm scale, SNeIa}$ become SNe Ia during every time step. The exponential law allows us to tie the amount of SNe Ia directly to the number of WDs produced from evolved AGBs in our model, while the power law DTD from \cite{2012MNRAS.426.3282M} and \cite{2017ApJ...848...25M} relates the amount of SNe Ia to the SFR.

The fraction of WDs that ultimately become SNe Ia is determined by $f_{\rm SNeIa}$ at the time of birth of AGB progenitors with initial masses between 3.2 and 8.5 M$_{\odot}$. In our fiducial model, $f_{\rm SNeIa}$ starts at 6.5\% for the first three Gyr to facilitate the rapid evolution in \alphafe{} in the beginning and then linearly decreases to 5\% at the last time step. This is inspired by \cite{2020MNRAS.499.1607M} who found that close binary fraction is higher at high \alphafe{}. We create four alternate models to explore the effect of SNe Ia. The first pair of models are named \textit{SNIa+} (a higher fraction of SNe Ia) and \textit{SNIa-} (a lower fraction of SNe Ia) as their $f_{SNeIa}$ is held constant at 6.5\% and 5\% respectively. The second pair of models are named \textit{DTD+} (a longer delay timescale) and \textit{DTD-} (a shorter delay timescale) and they adopt $t_{\rm scale, SNeIa}=2$ Gyr and $t_{\rm scale, SNeIa}=1$ Gyr respectively. 

\subsection{Infall} \label{subsec:infall}

Galaxies accrete their gas from the IGM in two distinct modes: the cold mode dominates low-mass galaxies and at high redshifts and the hot mode dominates high-mass galaxies and at low redshifts. However, the gas accreted from the cold mode is still at least $10^4$ K and would take a significant amount of time to cool before it can form stars. We experimented with depositing infalling gas into the warm gas components in our model instead of cold gas and found it difficult to reach a high enough SFR to drive chemical evolution effectively in the first 2 Gyr to match the age-\feh{} relation from \cite{2021MNRAS.507.5882S}. A possible improvement in future could be to vary the ratio of hot vs. cold accretion over time as recently quantified by \cite{2021MNRAS.504.5702W} from cosmological simulations.

We adopt an infall rate over time as prescribed by \cite{2009MNRAS.396..203S}, which is the sum of two exponential terms, 
\begin{equation} \label{eq:infall}
    \frac{dM}{dt} = \frac{M_1}{b_1}e^{-b_1/t} + \frac{M_2}{b_2}e^{-b_2/t}.
\end{equation}
Here, $M_1 = 4.0 \times 10^{10} \rm M_\odot$, $b_1 = 300$ Myr, $M_2 = 7 \times 10^{10} \rm M_\odot$, $b_2 = 14$ Gyr. These numbers are roughly chosen based on the following two constraints. First, the fiducial model needs to have about $5 \times 10^{10} \rm M_\odot$ in stellar mass at the last time step as constrained by the Milky Way's present-day properties \citep{doi:10.1146/annurev-astro-081915-023441}. Second, the SFR at the last time step needs to be approximately a few solar masses per year. Although our Equation \ref{eq:infall} has two exponential terms and looks similar to that of a classical two-infall model, there is no time delay between the terms. The long-timescale exponential term sustains star formation throughout the entire time span of our model but we found it necessary to include a short-timescale term at the beginning to match the density of the high-\alphafe{} sequence.

The mass of infalling gas during a time step is approximated by $M_{infall, t} = dM/dt \times \Delta t$. The final amount of cold gas in each zone by the end of radial inflow is obtained by multiplying its corresponding relative mass fraction ($f_{n, {\rm cold}}$) in Section \ref{subsec:setting up} by the sum of the mass of infalling gas ($M_{{\rm infall}}$) and the total mass of cold gas in the model ($\sum_i M_{i, {\rm cold}, t}$) during a given time step, or $M_{n, {\rm cold}, t}' = (M_{{\rm infall}, t} + \sum_i M_{i, {\rm cold}, t}) \times f_{n, {\rm cold}}$. The difference in cold gas mass in each zone, $\Delta M_{n, t} = M_{n, {\rm cold}, t}' - M_{n, {\rm cold}, t}$, will be gained through fresh gas directly from the IGM or an outer zone as a result of radial inflow or more commonly both. In an alternate model named \textit{CSF} (constant star formation), we explore an alternate scenario by fixing the amount of cold ISM to be $8.3 \times 10^9$ after infall at every time step so that the final stellar mass in the model is the same as in the fiducial model. 
 
The chemical composition of the infalling gas is pristine (only H, He, and Li) when the GCE starts. As the model evolves, the metallicity of the infalling gas matches that of the outermost zone until it reaches $\log_{10}{Z/Z_{\odot}} = -0.8$ with $Z_\odot = 0.0156$. Afterwards, the infalling gas continues to mirror the abundance composition of the cold gas in the outermost zone but its metallicity remains fixed (at $\log_{10}{Z/Z_{\odot}} = -0.8$). The solar metallicity and abundances in this work are taken from \cite{2009ARA&A..47..481A}. The infalling gas determines the metallicity and abundance pattern of the outer zones especially past 10 kpc and places constraints on the abundance pattern in the model because the fresh gas can dilute existing gas and slow down chemical enrichment.

\subsection{Gas dynamics} \label{subsec:gas}

All of our models have intra- and inter-zone mechanisms for gas dynamics. Star formation and supernovae have the potential to heat up cold ISM in our models. \cite{2014MNRAS.445..581H} showed in high-resolution cosmological simulations that supernova and star formation feedback both shape the SFH of galaxies. The amount of cold ISM heated up by star formation activities is $\eta_{\rm SF}$ multiplied by the amount of cold ISM involved in star formation. Similarly, $\eta_{\rm SN}$ is the multiplier for supernovae. Although the total mass loading factor can be calculated from high-resolution hydrodynamic simulations, it is difficult to isolate the effect of individual processes. \cite{2017ApJ...841..101L} and \cite{2020ApJ...900...61K} calculated the total mass loading factor to be between one and ten for conditions covered in our models. Since supernova-driven outflow is more dominant, we chose $\eta_{\rm SF}=1.0$ and $\eta_{\rm SN}=3.0$ at all radii for all models.

Within each zone, stellar winds from AGB and supernovae explosions release the processed and unprocessed metal trapped inside the stellar component into the surrounding gas. A fraction $f_{\rm direct}$ goes to the cold ISM and a fraction $f_{\rm eject}$ goes to the IGM which is considered as lost by the model. The rest ($1 - f_{\rm eject} - f_{\rm direct}$) of the produced metal goes into the warm ISM, which cools off exponentially over a time scale of $t_{\rm cool}=1.2$ Gyr and feeds enriched gas to the cold ISM. In the fiducial model, $f_{\rm direct}=0.01$ and $f_{\rm eject}=0.2$ at all radii.  These fractions determine where the newly produced elements end up immediately after they are produced and have a large impact on the rate of chemical evolution. \cite{2019ApJ...873..129P} estimated that only 30\% of produced metals are inside the galaxies. Most metals end up in the IGM after they are produced. We chose $f_{\rm direct}=0.01$ because some ejecta would have cooled within one time step that lasts 30 Myr. In practice, the effect of $f_{\rm direct}$ is negligible and essentially all the newly produced elements go through the warm ISM before they are recycled for future rounds of star formation. The effect of $f_{\rm direct}$ is highly degenerate with that of the cooling timescale because stellar evolution is one of two sources that feed the warm ISM. In two alternate models, \textit{E+} (higher ejection ratio) and \textit{E-} (lower ejection ratio), we explore the effect of $f_{\rm eject}$ by setting them to 0.4 and 0 respectively, as $f_{\rm eject}$ is expected to influence the rate of chemical enrichment. 

\begin{figure}[!ht]
  \includegraphics[width=0.9\linewidth]{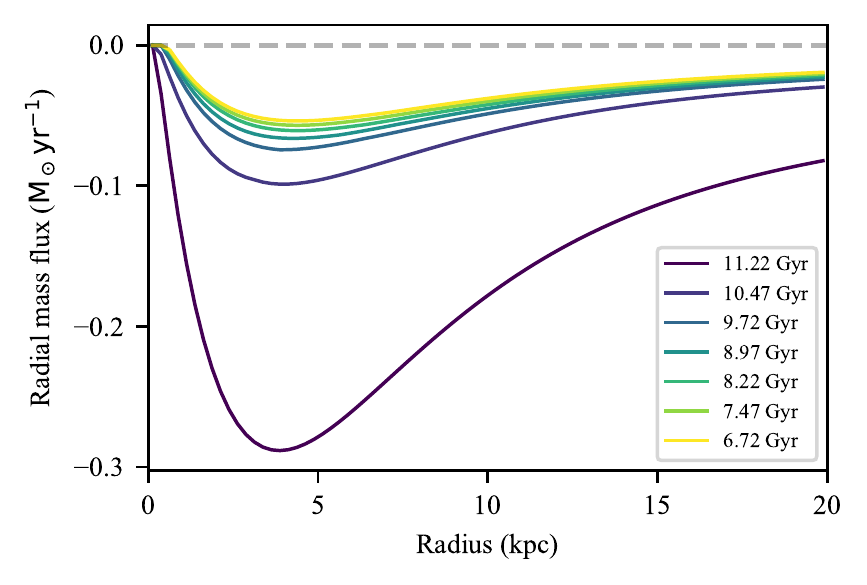}
 \vspace*{-3mm}
  \caption{Radial mass influx of radial inflow in the fiducial model. Each line corresponds to a specific lookback time from twelve to six Gyr ago. }
  \label{fig:mass_influx}
\end{figure}

We implement the radial flow of gas because it could be crucial to generating a radial chemical gradient. Our radial inflow scheme is inspired by \cite{2021arXiv210511472T} which shows in cosmological simulations the majority of fresh gas is carried from the outskirts into the inner regions of galaxies. This is also confirmed by observational evidence in \cite{2021ApJ...918L..16C} in an edge-on system. Our scheme is effectively similar to the one in \cite{2009MNRAS.396..203S} but it is more numerically stable when handling a large infall rate or a smaller time step. 

Fresh gas from the IGM joins the disk in our models in two steps. Firstly, a fraction $f_{\rm IGM}$ of fresh gas in a given time step falls into each zone in situ according to the pre-determined mass fractions in Equation \ref{eq:density} and \ref{eq:mass_frac} in Section \ref{subsec:setting up}. This corresponds to the ``raining down'' scenario of accreting IGM. Secondly, the rest of the infalling gas is deposited onto the outermost zone and carried inward so that exactly $\Delta M_{n, t}$ is left in each zone after radial inflow is completed. The inflowing gas is mixed instantaneously with the existing gas reservoir in each zone along its path. Since a significant amount of gas is accreted in the first two Gyr, the inflowing gas is rapidly enriched as it flows through the disk. We chose $f_{\rm IGM} = 0.3$ for our standard model which is sufficient to generate observed chemical gradients. Figure \ref{fig:mass_influx} shows the average radial mass influx induced by our radial inflow mechanism over circles with the central radii of our zones, which exhibits similar behaviours to model \textit{m12m} and \textit{m12f} in the lower panel of Figure 4 in \cite{2021arXiv210511472T}. A more sophisticated treatment of radial inflow is provided by \cite{2015A&A...580A.126K} who assume that the radial flow of gas is induced by the bar and gas from within the co-rotation radius ($\sim 3.5$ kpc) flows outwards instead of inwards (see their Figure 5). However, the gas flow profiles beyond 5 kpc in both schemes are similar. 

In low-redshift cosmological simulations, the speed of gas flow is typically a few km/s, which roughly translates to tens to a hundred Myr delay between two adjacent zones. When the infall rate is high in the first two Gyr, the speed of gas radial flow would be too fast because a large amount of fresh gas in the outer zones would spill over and travel inwards for several kpc before it settles down in one time step. Nevertheless, during the early phase, the metallicity of the cold ISM in the outer zones is similar to the infalling gas and there is minimal metallicity gradient in the disk at this time. Thus, the composition of the gas flowing through the disk barely changes after travelling for a long distance and can be treated as gas ``raining down'' instead. The infall rate is much lower during the low-\alphafe{} regime. Fresh gas only fills the outermost zones and the radial inflow primarily happens between adjacent zones. In an alternate model, \textit{MRI} (Minimum radial inflow), we set $f_{\rm IGM} = 1.0$ and only allow radial inflow to maintain the scalelength of the cold ISM.

\subsection{Radial migration} \label{subsec:churning}

We implement the churning mechanisms (change in guiding radius) from \cite{2009MNRAS.396..203S} and \cite{2015MNRAS.449.3479S}. The former ensures that an equal amount of cold gas and stars are migrating radially inwards and outwards by relating the churning probability at a radius to the amount of mass present in its corresponding radial zone and its closest neighbours. Since the same amount of mass migrates between two zones at any time, the angular momentum in the model is conserved. The amplitude of churning is controlled by a free parameter $k_{ch}$ which is the maximum churning probability at any radii. The latter is an analytic prescription from solving the action-space diffusion equation by assuming an exponential disk with a fixed scalelength $R_d$ and constant rotational speed, which also conserves the angular momentum. The churning amplitude of this mechanism is constrained by observational data. We used Midpoint Rule to approximate the integral of Equation 23 in \cite{2015MNRAS.449.3479S} over each zone and used the same parameters as those in their Figure 3. To be clear, we assume a constant rotation curve with a speed of $v_0=220$ km/s, the birth of the Galaxy is 12 Gyr ago, and a churning amplitude  $\sigma_{L_0}=1150$ kpc km/s which controls the strength of radial migration. We briefly describe the first churning mechanism here to avoid confusion because we are unable to match Figure 3 from \cite{2009MNRAS.396..203S} with the same $k_{ch} = 0.25$. 

\begin{figure}[!h]
  \includegraphics[width=\linewidth]{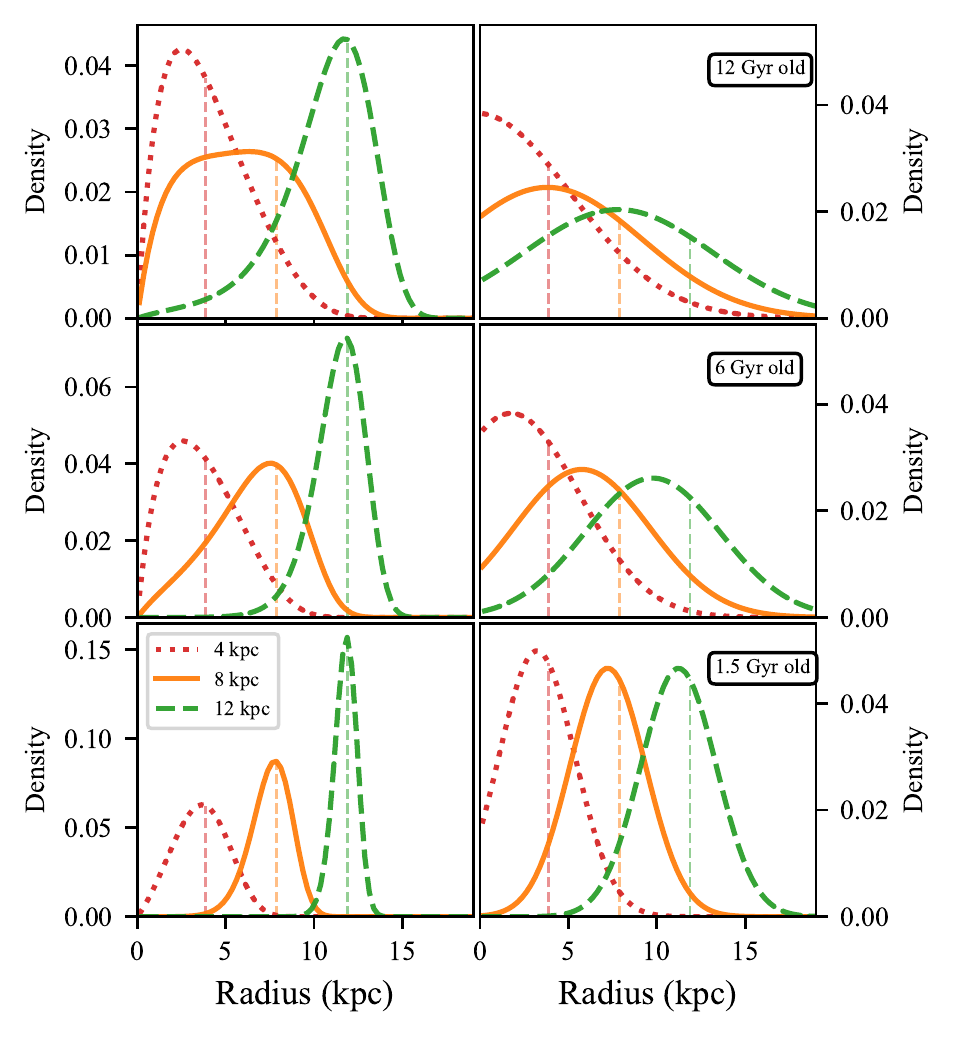}
  \caption{The distribution of the guiding radii of stars born at three zones closest to 4, 8, and 13 kpc (in red, orange and green and highlighted with vertical lines) with ages of 12, 6, and 1.5 Gyr from top to bottom panel. The left column shows the distribution with churning mechanism from \cite{2009MNRAS.396..203S} ($k_{ch}=0.25$) and the right column from \cite{2015MNRAS.449.3479S}. }
  \label{fig:churning_prob}
\end{figure}

Section 2.5 in \cite{2009MNRAS.396..203S} provided justification for basing the churning probability on mass. The probability for cold gas and stars in zone $i$ to be transferred to zone $j$ is defined as 
\begin{align} \label{eq:churning}
    p_{ij} = 
    \begin{cases}
            k_{ch} M_j / M_{\rm max} & \mathrm{for \ j  = i \pm 1}\\
           0 & o.w.
    \end{cases}
\end{align}
\noindent where $M_{max} = max_j(M_j)$ is the maximum mass of any radial zone. The amount of mass transfer between two adjacent zones $i$ and $i+1$ is the same as $M_{i, i+1} = p_{i, i+1} M_{i} = (k_{ch} M_{i+1} / M_{max}) M_{i} = (k_{ch} M_{i} / M_{max}) M_{i+1} = p_{i+1, i} M_{i+1} = M_{i+1, i}$ and thus the total angular momentum in the model is conserved. When $M_j = M_{max}$, $k_{ch}$ is the churning probability for the most massive zone. Similar to \cite{2009MNRAS.396..203S}, we allow two churning operations per time step to allow each zone to exchange materials with the second nearest zone but recalculate the churning matrix at every time step. 

The effect of churning by these two mechanisms is shown in the left and right columns respectively in Figure \ref{fig:churning_prob}. We picked stellar populations that are 1.5, 6, and 12 Gyr old at three representative radii, 4, 8, and 13 kpc. The stars at 4 kpc from both mechanisms experience a significant amount of scattering as the peak guiding radius shifts away from its birth radius. They can reach as far as 7 kpc after just 1.5 Gyr. The effect of churning starts to differ significantly as we move away from the inner zones. In the left column, stars born at 13 kpc are far less likely to migrate and stars born at 8 kpc 12 Gyr ago are mostly limited to within 10 kpc, breaking the bell shape. Stars born at 4 kpc in the left column are reluctant to migrate towards the galactic centre and tend to be stuck around 3 kpc, while the majority of the same stars born at 4 kpc in the right column end up at the galactic centre. 

These behaviours are caused by the exponential density of the disk and the prescription for radial migration by \cite{2009MNRAS.396..203S}. The churning probability in Equation \ref{eq:churning} is related to the amount of mass in each zone. Due to the exponential density profile, there is little mass in the outskirts of our model and thus it is far less likely for stars to migrate there. Similarly, the area of a ring diminishes as we move towards the centre, bringing the mass and churning probability ultimately to zero. Rigorously, we can differentiate the formula that describes the amount of mass in each radial zone to find out which radial zone has the most amount of mass and attracts the most stars.
\begin{align} \label{eq:mass}
    M_R &\sim e^{-R/R_d} \times 4\pi R\delta R \sim e^{-R/R_d}R \\
    dM_R/dR &\sim (e^{-R/R_d}R)' = (1 - R/R_d) e^{-R/R_d}
\end{align}
When $R=R_d$, $dM_R/dR=0$ and the maximum mass in any zone is reached. If we move away from the radial zone at $R_d$, the proportion of mass in each zone decreases and so does the churning probability. Since we are interested in radial zones away from the solar neighbourhood, we adopt \cite{2015MNRAS.449.3479S} as the churning mechanism for our standard model to avoid stars being stuck at their birth radii. Nevertheless, we show the effect of the churning mechanism from \cite{2009MNRAS.396..203S} in an alternate model named \textit{C09}.

\begin{figure}[!h]
  \includegraphics[width=\linewidth]{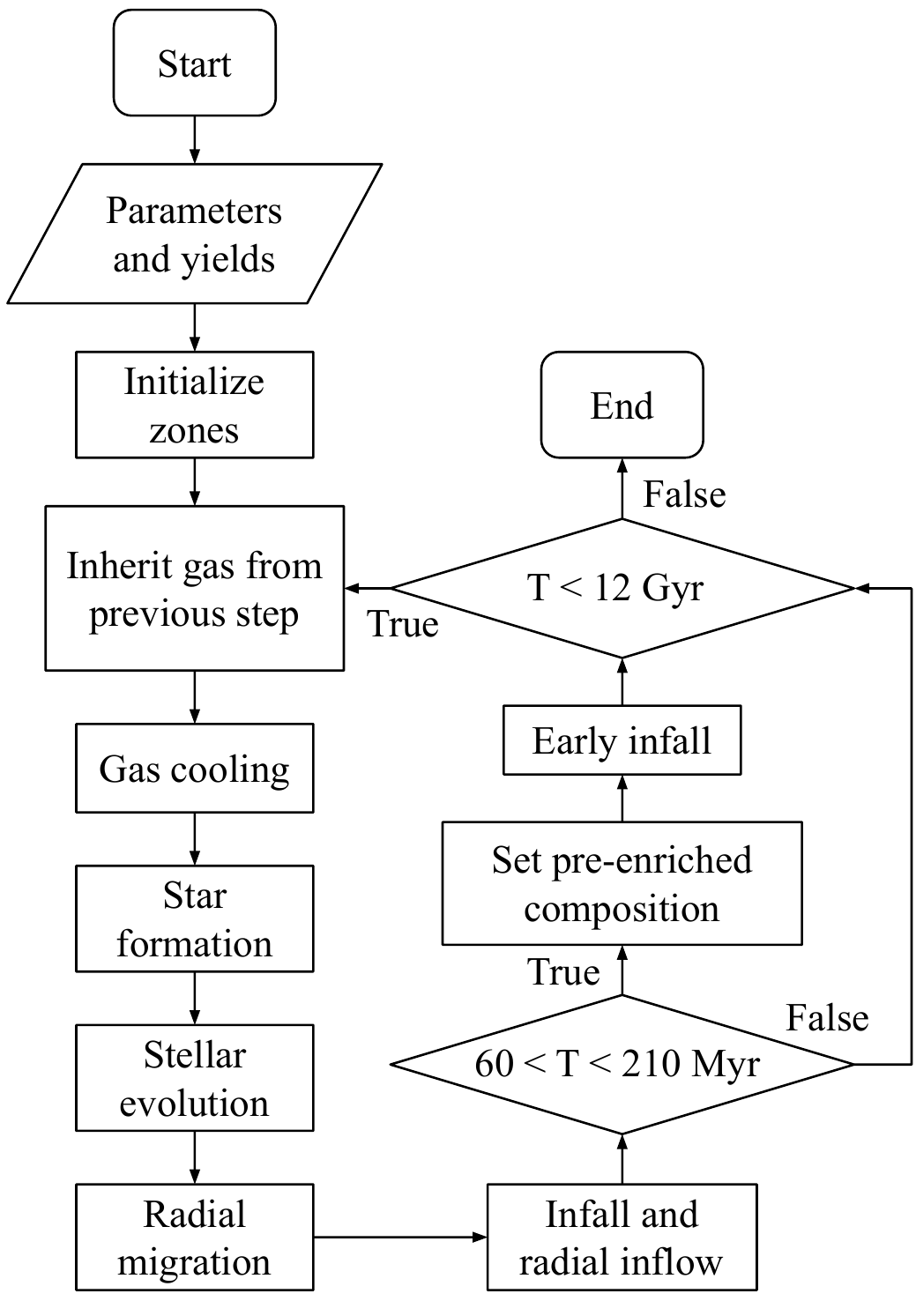}
  \vspace*{-7mm}
  \caption{Flowchart that demonstrates the workflow of the galactic chemical evolution model in this work.}
  \label{fig:flowchart}
\end{figure}

\subsection{Present day phase space distribution} \label{subsec:phase_space}
The chemical evolution model gives the distribution of newly formed stars for a given age $\tau$ and birth radius $R_b$. To make observational predictions using our chemical evolution tracks, we need to generate the present-day phase space distribution $p({\bf x},{\bf v})$ of stars where $\bf x$ is the three-dimensional position and $\bf v$ is the three-dimensional velocity. We use the same prescription as \citet{2021MNRAS.507.5882S} which samples stars from a phase space distribution that is in equilibrium in a 2D gravitational potential in the $R$ and $z$ direction. The full distribution is given by 
\be
p({\bf x},{\bf v})=p({\bf x},{\bf v}|R_b,\tau, L) p(L|\tau,R_b) p(\tau,R_b)
\ee
The scheme uses radial migration from \citet{2015MNRAS.449.3479S} 
to generate the distribution of angular momentum $p(L|R_b,\tau)$ for a given age $\tau$ and birth radius $R_b$. The distribution $p({\bf x},{\bf v}|R_b,\tau,L)$  is of the following form \citep[see Equation 4.147 from][]{2008gady.book.....B}.  
\be
f(E_R,L,E_z) \propto \frac{F(L)}{\sigma_R^2}\exp\left(-\frac{E_R}{2\sigma_R^2}\right)\exp\left(-\frac{E_z}{2\sigma_z^2}\right)\, 
\ee
where $L$ is specific angular momentum,  
$E_R$ is radial energy and $E_z$ is vertical energy which is a function of velocity and Gravitational potential 
$\Phi(R,z)$.  The planar part is modelled by the Shu distribution while the vertical part is modelled by an isothermal distribution. It then uses \citet{2021MNRAS.506.1761S} to get the velocity dispersions $\sigma_{v_R}$ and $\sigma_{v_z}$ as a function of $L$, $\tau$, and $R_b$. For the gravitational potential $\Phi(R,z)$ it adopts MWPotential2014 from galpy \citet{2015ApJS..216...29B}.

\subsection{Summary} \label{subsec:model_summary}
We presented all the ingredients present in our model in this section. We summarize all the parameter values employed by our fiducial model in Table \ref{table:parameters}. We also created eleven other models to explore the effect of radial migration, radial inflow, SFH and SNe Ia DTD. The changes made to these alternate models are summarized in Table \ref{table:alternate}. These models explore a subset of parameters that have the most effect on the chemical evolution of the Milky Way. Lastly, we are able to extend our 1D model into a 2D model along the z height by making assumptions about the potential of the Milky Way. The sequence of events in all of our models is summarized in a flowchart in Figure \ref{fig:flowchart}. 

\begin{figure*}
  \includegraphics[width=0.8\linewidth]{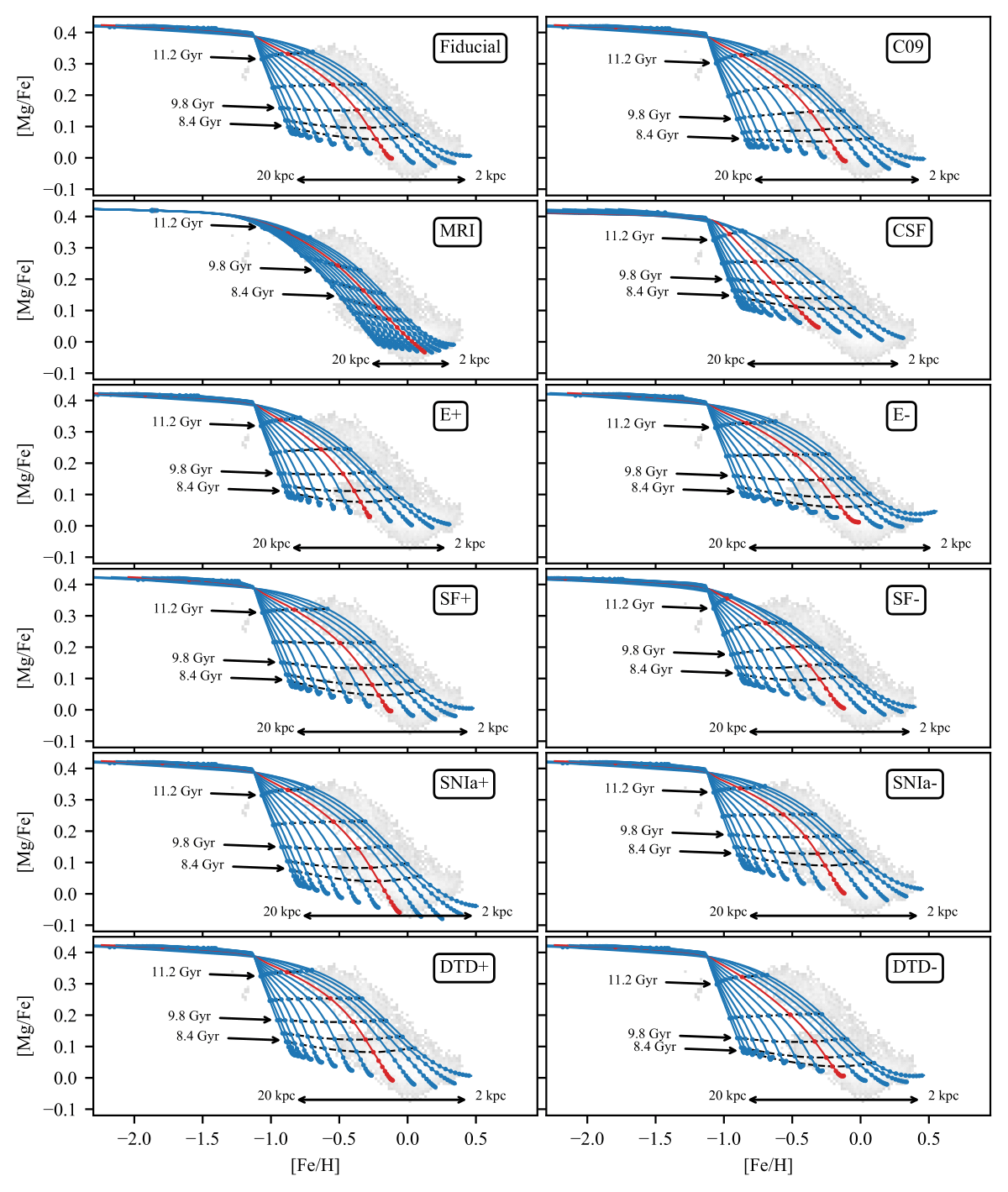}
  \centering
  \vspace*{-3mm}
  \caption{Chemical evolutionary tracks for GCE models. The fiducial model and eleven other models summarized in Table \ref{table:parameters} and Table \ref{table:alternate} are shown here. In each panel, the tracks are shown in 1.5 kpc increments, with the track closest to 8 kpc highlighted in red. Abundances belonging to different radial zones at the same time step are connected every 750 Myr (25 time steps) until 8.4 Gyr ago. }
  \label{fig:tracks}
\end{figure*}

\section{Results} \label{sec:results}

\subsection{Comparison with observational data} \label{subsec:data}
As one of our primary goals in this work is to explain the spatial variation of the thin and thick disks shown in \cite{2015ApJ...808..132H}, we will compare the results from our models to APOGEE DR16 \citep{2020ApJS..249....3A}, which due to 
its use of infrared wavelengths provides one of the best radial coverage of the Galaxy. We implement the following cuts to restrict our sample to giants: $1.0 < \log\ g < 3.5$, $3500 < T_{\rm eff} < 5300$ and $7 < H < 11$. We will compare the 2D joint MDF in the \fehalpha{} plane in different (R, z) bins in galactocentric coordinates to those simulated in our models according to the phase space distribution specified in Section \ref{subsec:phase_space}, even though our models only contain radial zones. When we compare our results in the solar neighbourhood, we restrict our sample to $7.5<R<8.5$ kpc, $|z|<2$ kpc. However, when we study the evolution of \alphafe{} over time, we opt for the main-sequence turn-off (MSTO) stars in GALAH DR3 \citep{2021MNRAS.tmp.1259B} by applying the following cuts: $3.2 < \log\ g < 4.1$, $5000 < T_{\rm eff} < 6100$ and $\mathrm{SNR} > 10$. The GALAH MSTO stars typically have an age precision of around one Gyr. \citet{2019MNRAS.489..176M} derived high-precision stellar ages for 65,719 stars in APOGEE and later extended their method to all APOGEE stars, but the age of their high-\alphafe{} sequence is only about 8-9 Gyr, while it is more than 10 Gyr old according to GALAH MSTO stars. Their results are limited by their training set from APOKASC, a common set of stars between \textit{Kepler} and APOGEE, which {is restricted to a narrow range of stellar parameters}. An earlier study by \cite{2018MNRAS.475.5487S} on 1979 APOKASC stars placed most high-\alphafe{} stars around 10-12 Gyr old, consistent with our GALAH MSTO sample. We opt to only use the ages of GALAH MSTO stars in order to compare our results to those from \citet{2021MNRAS.tmp.1920S}. 

\begin{figure*}
  \includegraphics[width=0.8\linewidth]{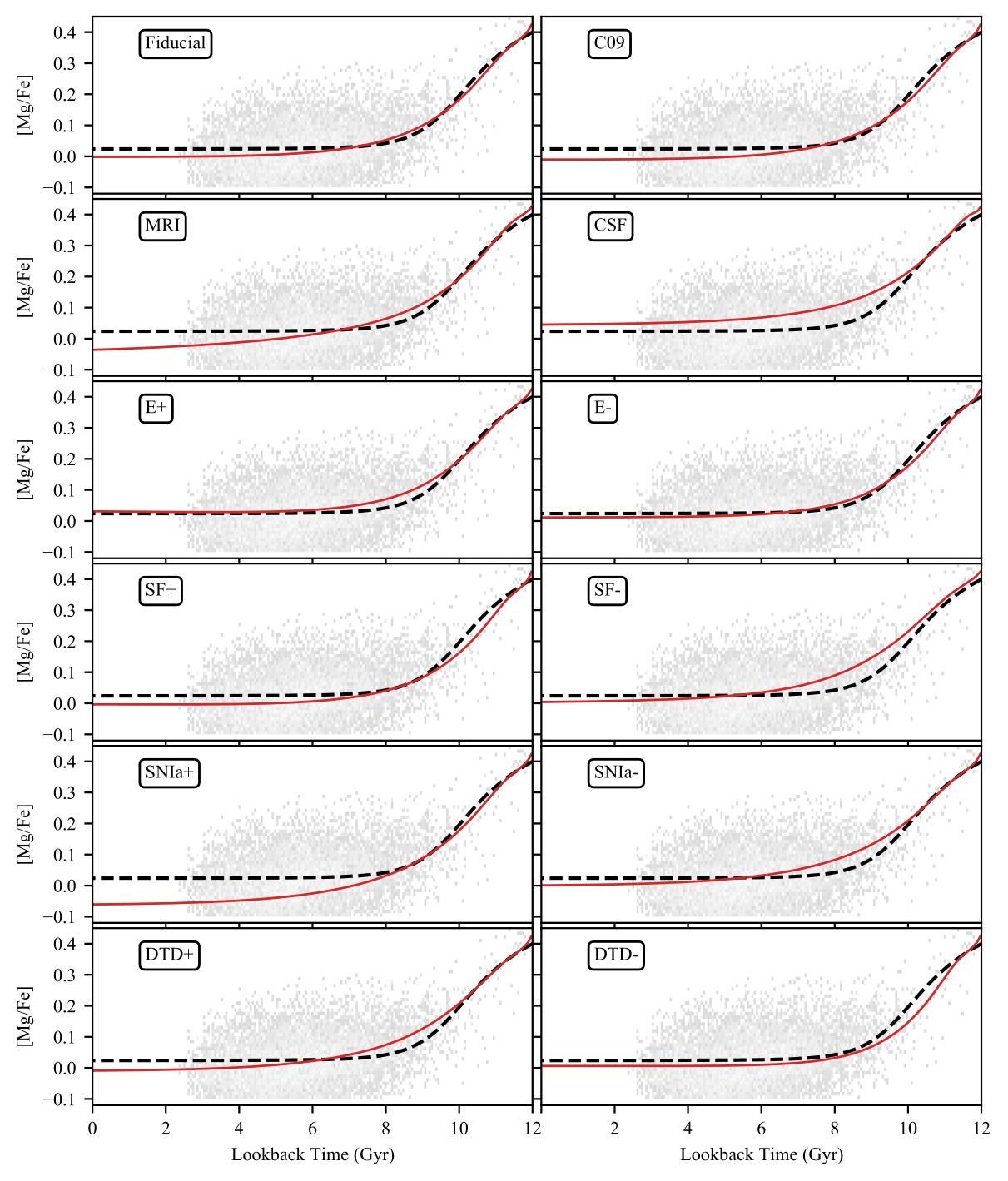}
  \centering
  \vspace*{-3mm}
  \caption{Evolution of [Mg/Fe] over time as predicted by our twelve GCE models. The grey histogram background is from the MSTO sample in GALAH DR3 whose stellar ages are trained on asteroseismic data. The dashed line is predicted by the analytic model extracted by APOGEE DR16 from \cite{2021MNRAS.tmp.1920S}. Both the dashed line and histogram are scaled to start from 0.4. The red lines are what our models predict at a 8 kpc.}
  \label{fig:alpha_vs_time}
\end{figure*}

\begin{figure*}
  \includegraphics[width=0.8\linewidth]{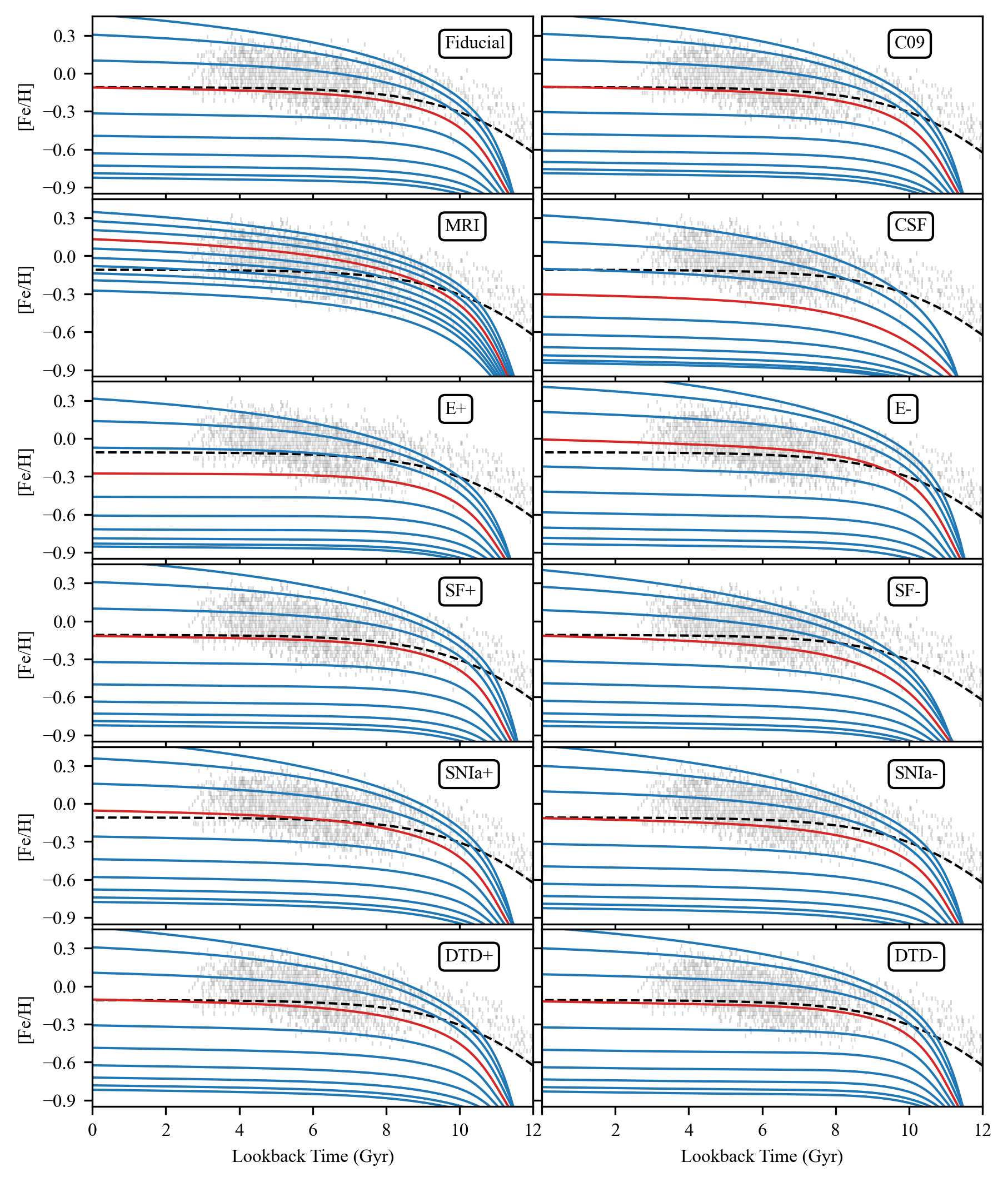}
  \centering
  \vspace*{-3mm} 
  \caption{Evolution of [Fe/H] over time as predicted by our twelve GCE models. The grey histogram background is from the MSTO sample in GALAH DR3 whose stellar ages are trained on asteroseismic data. The dashed line is predicted by the analytic model extracted by APOGEE DR16 from \cite{2021MNRAS.tmp.1920S}. Here we show tracks from 2 kpc to 20 kpc spaced 2 kpc apart with the 8 kpc track highlighted in red to demonstrate the evolution of radial gradient for \feh{} over time.}
  \label{fig:fe_vs_time}
\end{figure*}

For each of our models, we created five diagnostic plots to showcase the effect of the highlighted parameters and compare our models to observational data. The first is the chemical evolutionary tracks in [Mg/Fe] vs [Fe/H] generated by our models compared to APOGEE data in the solar neighbourhood, shown in Figure \ref{fig:tracks}. The tracks are shown in 1.5 kpc increments and the track belonging to the zone with a central radius closest to 8 kpc is highlighted in red. A dashed line connects the abundances of all radii shown in a given time step every 750 Myr, providing a reference guide in evolution time and the radial gradient of [Mg/Fe]. Three representative time steps are highlighted in each panel with lookback times of 11.2, 9.8, 8.4 Gyr. SNe Ia started to kick off and influence the chemical evolution about 11.5 Gyr ago in our models. Since the infalling gas mirrors the composition of the existing cold ISM in the outermost zone, the drop in \alphafe{} in the ISM is also reflected in the infalling gas. A large amount of infalling gas with different \alphafe{} in the first Gyr causes a sharp turn in the chemical tracks in \fehalpha{}, especially in the outer zones. This discontinuity could be solved by changing the temperature mode of the infalling gas from cold to warm in the future so that the warm ISM acts as a buffer between infall and the existing ISM but it would require further refinement of the model. By 9.8 Gyr ago, [Mg/Fe] in our models reached the highest value observed for the stars in the low-\alphafe{} sequence in the solar neighbourhood. [Mg/Fe] approached equilibrium values 8.4 Gyr ago at outer radii, while zones less than 10 kpc away from the galactic centre continued to evolve gradually for the remainder of the time. The bottom of each panel shows the minimum and maximum radii of tracks in the model from 2 to 20 kpc. 

\begin{figure*}
  \includegraphics[width=0.8\linewidth]{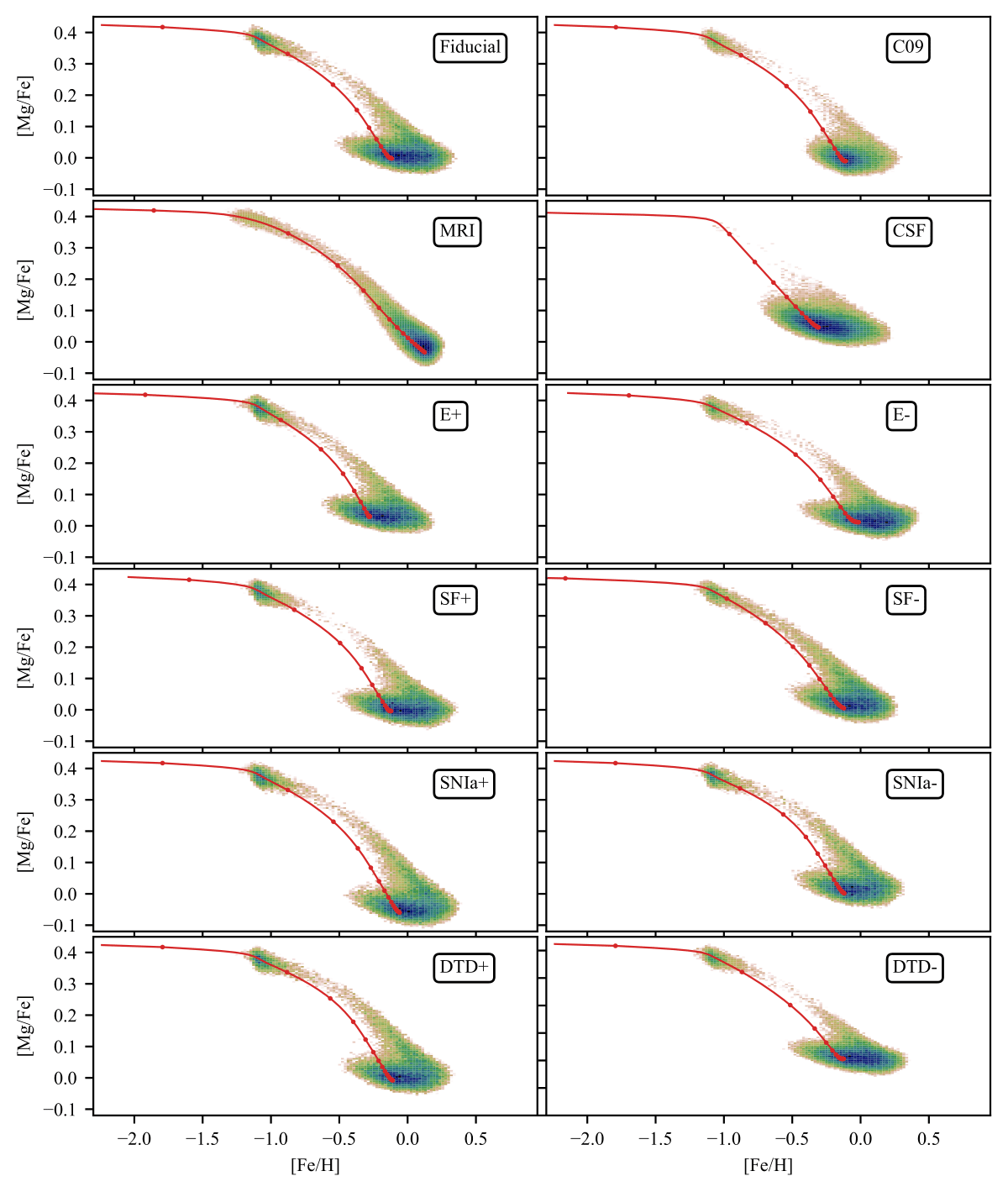}
  \centering
  \vspace*{-3mm}
  \caption{2D histogram in [Mg/Fe] vs [Fe/H] as predicted by our twelve GCE models for the solar neighbourhood. We select the zone at 8 kpc which is 0.25 kpc wide to represent the solar neighbourhood. The chemical track for this zone in each model is shown in red.  }
  \label{fig:hist_2d}
\end{figure*}

The second figure is the evolution of [Mg/Fe] over time in comparison to the GALAH DR3 MSTO sample. Although Figure \ref{fig:tracks} can provide some general sense of how fast [Mg/Fe] evolves over time in our models, Figure \ref{fig:alpha_vs_time} offers a direct comparison to observational data and the analytic results from \cite{2021MNRAS.507.5882S}. The rapid evolution  [Mg/Fe] in our models always starts at 0.4 dex as dictated by the CCSN yield, which is higher than the value (0.3 dex) observed for the high-\alphafe{} stars in APOGEE or GALAH. Therefore, we had to scale \alphafe{} in APOGEE and GALAH such that their highest \alphafe{} values are also 0.4 for direct comparison. The grey histogram in the background shows the distribution of GALAH MSTO stars in [Mg/Fe] vs stellar age. The dashed line shows the analytic evolutionary track of \alphafe{} extracted from observational data in \cite{2021MNRAS.507.5882S}. Both the red and black dashed tracks in each panel correspond to a galactocentric distance of 8 kpc. The third is the evolution of [Fe/H] over time shown in Figure \ref{fig:fe_vs_time} in the same style as Figure \ref{fig:alpha_vs_time}. The only difference is that we show tracks in addition to the 8 kpc track every 2 kpc starting from 2 kpc because unlike [Mg/Fe] there is a significant radial gradient in [Fe/H]. The stars that deviate from the solar track come from other radial zones according to our models. The starting \feh{} of the analytic model and the \feh{} value of the oldest stars in the GALAH MSTO sample are both around -0.7, while our GCE models have to start from pristine gas. This causes the age-metallicity relations from our models to significantly differ from that observed in the GALAH data for the old stars. We have experimented with different parameter values and found it difficult to replicate the extremely rapid rise of \feh{} in a few time steps. As a result, the replicated high-\alphafe{} sequences in our models are more metal-poor than GALAH or APOGEE. The conditions immediately after the birth of the Milky Way require further study in the future.

\begin{figure*}
  \includegraphics[width=0.8\linewidth]{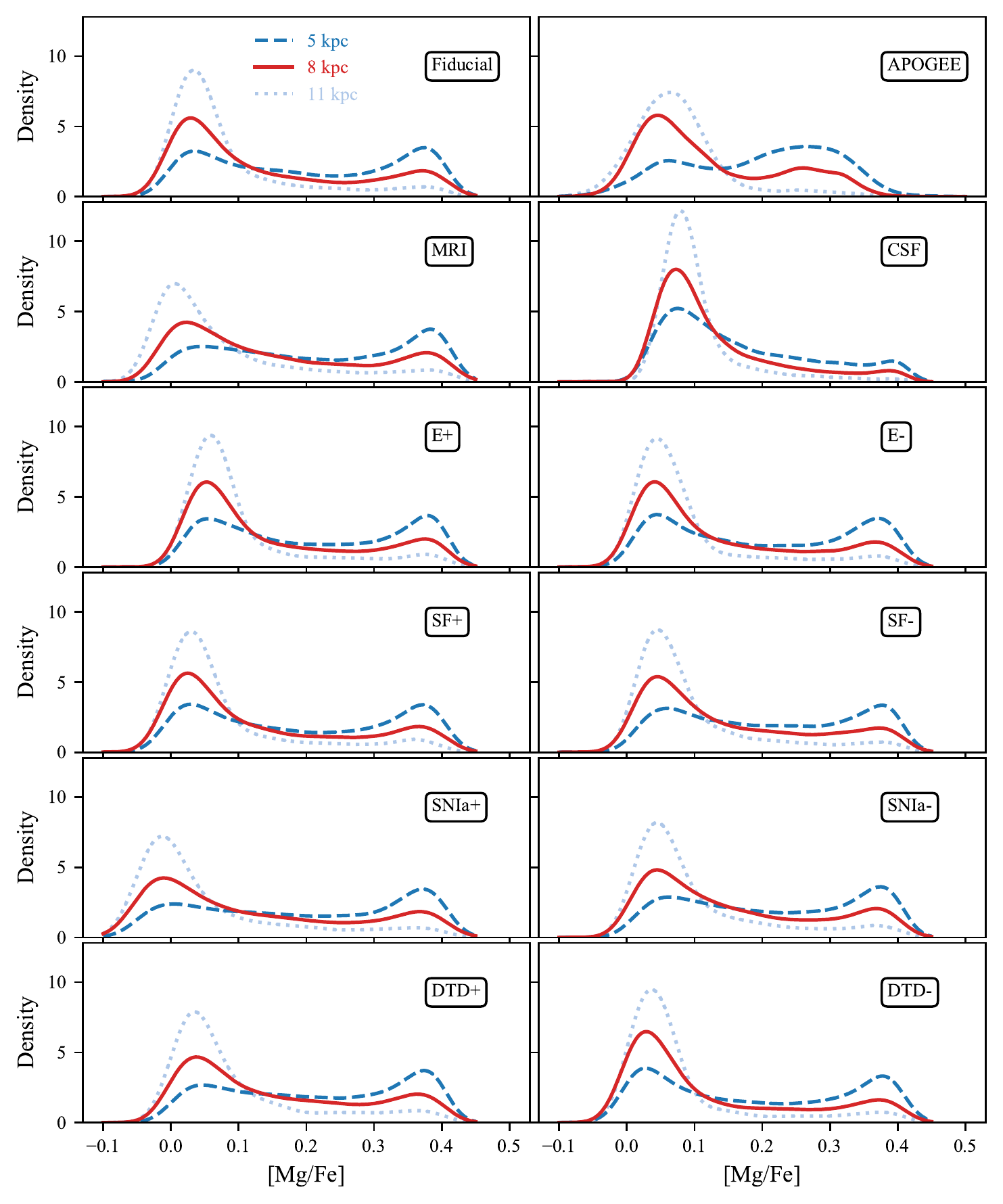}
  \centering
  \vspace*{-3mm}
  \caption{Distribution of [Mg/Fe] as predicted by our eleven GCE models (C09 omitted) and the APOGEE data at 5 (blue dashed), 8 (red solid), and 11 (light blue dotted) kpc with a width of 0.25 kpc. We applied a spatial selection criterion $500 < |z| < 1000$ pc to our models and the APOGEE data to minimize the effect of APOGEE target selection and limit our comparison to disk stars.  }
  \label{fig:alpha_hist}
\end{figure*}

The fourth is the density distribution in [Mg/Fe] vs [Fe/H] as predicted by our models shown in Figure \ref{fig:hist_2d}, incorporating the SFH and the effect of radial migration. The bins and density thresholds are the same in every panel. As the tracks can vary slightly for each model, we included the track whose central radius is the closest to 8 kpc in each panel in red as a reference. These tracks are already shown in the same colour in Figure \ref{fig:tracks}. The stars that deviate from the red tracks are present in the solar neighbourhood as a result of radial migration. We added observational uncertainty of 0.019 and 0.026 to simulated [Fe/H] and [Mg/Fe] respectively, which are equivalent to twice the median uncertainty (for intrinsic spread and observational uncertainty) reported in APOGEE DR16 for the solar neighbourhood sample. The last is a series of one-dimensional histograms in [Mg/Fe] for eleven out of twelve models and the APOGEE data at three different radii, 5, 8, and 11 kpc, shown in Figure \ref{fig:alpha_hist}. We omitted C09 here because it is indistinguishable from the fiducial model due to the lack of a substantial \alphafe{}-gradient in the disk. Although the infrared wavelength coverage of APOGEE can better penetrate the dust that obscures our Galaxy, it is still difficult to observe stars close to the Galactic plane away from the solar neighbourhood. Thus, we selected stars with $500 < |z| < 1000$ pc in our models and APOGEE data to minimize the effect of APOGEE target selection and limit our comparison to disk stars. The high-\alphafe{} becomes more dominant as we move closer to the Galactic center. The SFR is proportionally higher in the inner zones because of the radial exponential density profile and a large amount of high-\alphafe{} stars can form in these regions and are visible even away from the Galactic plane. 

\subsection{Radial migration} \label{result:radial migration}
Comparing model C09 to the fiducial model, radial migration has little effect on chemical evolution. The evolutionary tracks in Figure \ref{fig:tracks} and the evolution of \feh{} and \alphafe{} in Figure \ref{fig:alpha_vs_time} and \ref{fig:fe_vs_time} between the two models are nearly identical, even though stars that migrated from their birth radii contribute to the chemical evolution of their residing radial zones. The only difference is that the \alphafe{} gradient in the outer radial zones appears to be flat in C09, while the same region in the fiducial model shows a small gradient. We found that this difference is caused by the radial migration of gas as the churning prescription from \cite{2009MNRAS.396..203S} is applied to both stars and gas. Even though we demonstrated in Section \ref{subsec:churning} that the effect of churning at outer radii is minimal in C09, the additional mixing of gas was sufficient to erase the \alphafe{} gradient as there is little gas in the outer zones. This difference disappeared once we removed the churning effect on gas. 

The effect of churning is the most prominent in Figure \ref{fig:hist_2d} where we show the 2D distribution in \feh{} and \alphafe{}. As we explained in Section \ref{subsec:churning}, the churning probability is the highest at $R_d = 3.5$ kpc for C09 and diminishes as we move farther away. Stars in the outer zones and close to the Galactic centre are not able to migrate far away from their birth radii, even after six Gyr. It is evident from Figure \ref{fig:churning_prob} that the churning prescription from \cite{2015MNRAS.449.3479S} is stronger than C09, especially for young stars. It is unsurprising that the spread in \feh{}, especially on the metal-poor end, is much smaller in C09 than in the fiducial model. The densest region in \fehalpha{} in C09 is determined by the solar track in red, but a significant amount of stars in the fiducial region deviate from the solar track. The churning prescription is essential to replicating the observed chemical distribution in our Galaxy.

\subsection{Radial inflow} \label{result:radial_inflow}

The effect of radial inflow can be observed in every single one of our diagnostic plots between model MRI and the fiducial model. Most notably in Figure \ref{fig:tracks}, the radial gradient in \feh{} for model MRI is significantly smaller than any other model. Therefore, radial inflow is crucial to generating metallicity gradients in our models. With minimum radial inflow, all radial zones retain the newly produced metals from nucleosynthesis and receive fresh infalling gas in situ. Thus, the abundance patterns in each zone are only caused by differential gas density and infall rate in the model. Even though the vast majority of zones are more metal-rich in MRI than the fiducial model, the innermost zones in MRI are more metal-poor. The equilibrium abundance levels reflect the balance between nucleosynthesis and fresh infall. In the fiducial model, as metal-poor fresh gas is carried inwards from the outermost zone, it displaces a large portion of pre-enriched gas in outer zones and moves the newly produced metal inwards. The inflowing gas becomes enriched in this process until its metallicity matches one of the zones along its path. All of the zones past this point will be getting gas with metallicity similar to its existing gas from the inflow in the fiducial model, while these same zones will only receive metal-poor fresh gas in the absence of radial inflow. {Therefore, the absence of radial inflow makes the innermost zones more metal-poor and the outer zones more metal-rich.} Not surprisingly, with the absence of metallicity variation in the evolutionary tracks, we find minimal spread in \feh{} in the \fehalpha{} plane in Figure \ref{fig:hist_2d}, as only gas density is responsible for generating any chemical gradient. However, since there is a minimal \alphafe{} gradient in our models, the [Mg/Fe]-dichotomy of MRI is similar to that observed in the fiducial model in Figure \ref{fig:alpha_hist}. 

\subsection{Supernovae ejection} \label{result:ejection}

Supernovae ejection slows down the rate of chemical evolution and influences the final equilibrium abundance values by removing newly produced metal from the model. In Figure \ref{fig:tracks}, E- shows clear flattening of \alphafe{} as its tracks extend vertically to the right in \feh{} in the last several Gyr, while such a trend is far less visible for E+. The effect of supernovae ejection on individual abundances of \alphafe{} and \feh{} is more apparent in Figure \ref{fig:alpha_vs_time} and \ref{fig:fe_vs_time}. \alphafe{} for E- drops slightly faster than the analytic model suggests, while E+ lags behind the analytic values. The \feh{} level for the solar neighbour in E+ is also shifted at least 0.2 dex slower than E- due to the loss of metals. Because of the slower chemical evolution, the gap between the high- and low-\alphafe{} sequences is filled in for E+, causing a less prominent \alphafe{}-dichotomy in Figure \ref{fig:hist_2d} and \ref{fig:alpha_hist}. The original prescription in \cite{2009MNRAS.396..203S} assigns a higher ejection ratio for the inner disk with a galactocentric distance of less than 3 kpc to suppress star formation in the bulge and prevent the inner zones from becoming too metal-rich.

\subsection{Star formation history} \label{result:sfh}

\begin{figure}[h]
  \includegraphics[width=0.85\linewidth]{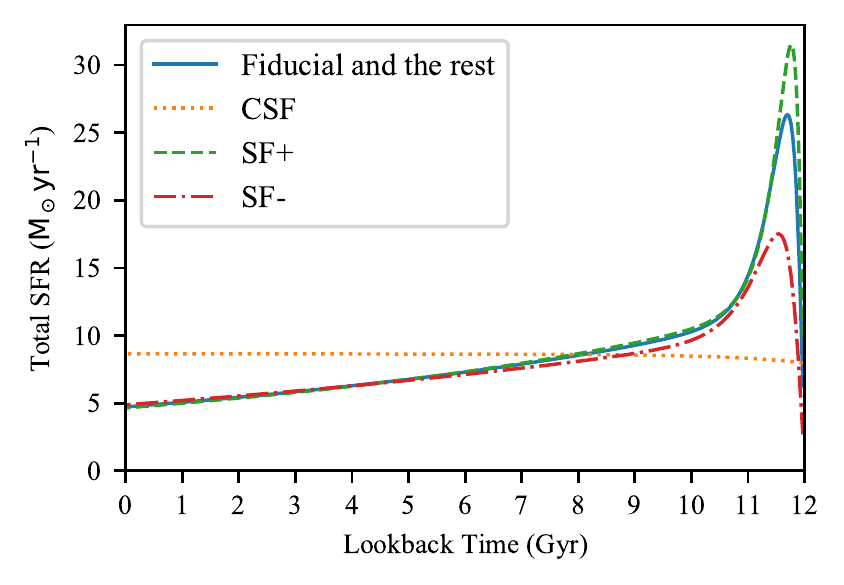}
  \centering
  \vspace*{-3mm}
  \caption{Average star formation history for our twelve GCE models. }
  \label{fig:sfr}
\end{figure}

Besides the fiducial model, we have three models with modified star formation prescriptions. The rest of the models share the same prescription as the fiducial model. The four distinct SFHs among our twelve models are shown in Figure \ref{fig:sfr}. In all of the star formation scenarios, the SFHs are nearly flat for the past 10 Gyr. As star formation drives nucleosynthesis, it is expected that SFH will have a tremendous effect on chemical evolution. Firstly, when we examine the chemical evolutionary tracks in Figure \ref{fig:tracks}, we find that CSF always has lower \feh{} and higher \alphafe{} than the fiducial model at every time step, even though both models produce the same amount of stellar mass in the end. Therefore, a peak in star formation rate during the early phase of the Milky Way is essential to driving the chemical evolution towards the observed values. Although the tracks from SF+, SF-, and the fiducial model appear similar in Figure \ref{fig:tracks}, a closer look at Figure \ref{fig:alpha_vs_time} and \ref{fig:fe_vs_time} reveals some subtle differences. All three models arrive at identical equilibrium values for \alphafe{} and \feh{}, but they reach there at different paces, with SF+ being the quickest and SF- the slowest. Therefore, the SFE has no effect on the equilibrium value of abundance levels but determines the rate of chemical evolution. 

The SFH also substantially affects the density distribution in chemical space. In Figure \ref{fig:hist_2d}, with the absence of an early peak in star formation, the high-\alphafe{} sequence in CSF is much less visible than the rest of the models because the SFR is not sufficiently high to build a substantial high-\alphafe{} sequence. As for SF+ and SF-, the SFE determines the prominence of the gap and the relative density between the high- and low-\alphafe{} sequences. When the SFE is higher, the model exhausts gas early when \alphafe{} is still high and forms fewer stars in the intermediate-\alphafe{} region as \alphafe{} falls rapidly. When the SFE is low, the model still has a considerable amount of gas in reserve as \alphafe{} drops and the \alphafe{}-gap is filled in by a considerable amount of star formation. The same behaviour can be observed in Figure \ref{fig:alpha_hist} where the median \alphafe{} of the low-\alphafe{} sequence for SF- is higher as SF- takes longer to reach equilibrium and more stars are formed at higher \alphafe{} as a result. 

\begin{figure*}
\centering
  \includegraphics[width=0.95\linewidth]{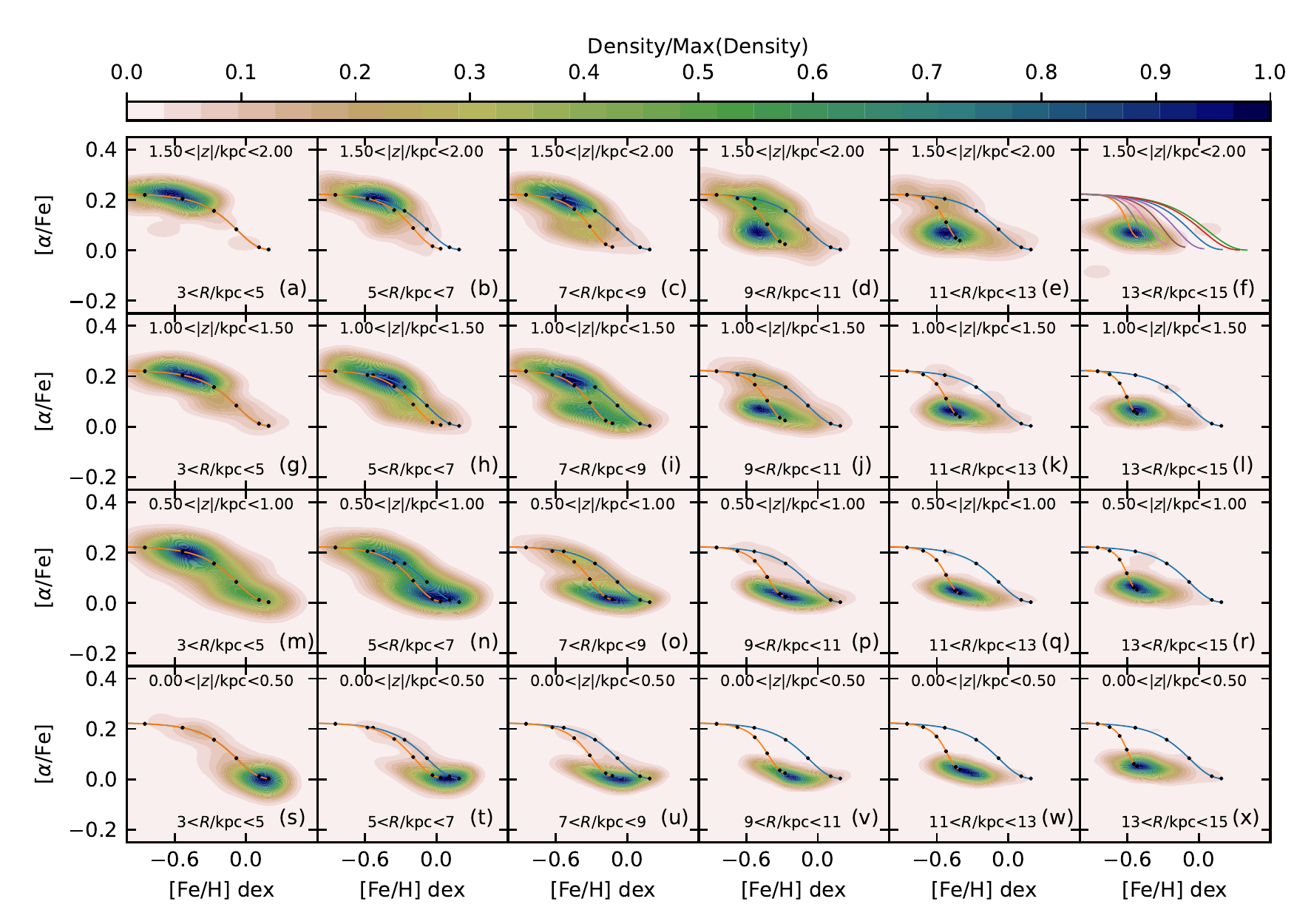}
  \vspace*{-2mm}
  \caption{The observed distribution of stars in the $([\alpha/{\rm Fe}], [{\rm Fe/H}])$ plane at different galactic (R,z) locations in cylindrical coordinates of APOGEE from \cite{2021MNRAS.507.5882S}. The best-fit analytic tracks for each galactocentric radius range from \cite{2021MNRAS.507.5882S} are shown in orange in each panel. The blue tracks belong to the region with $3 < R < 5$ kpc. The top right panel shows all the tracks extracted by the analytic model for different radii from APOGEE, starting with $R==1$ kpc, with a spacing of 2 kpc for each successive track.}
  \label{fig:rz_var_apogee}
\end{figure*}

\begin{figure*}
\centering
  \includegraphics[width=0.95\linewidth]{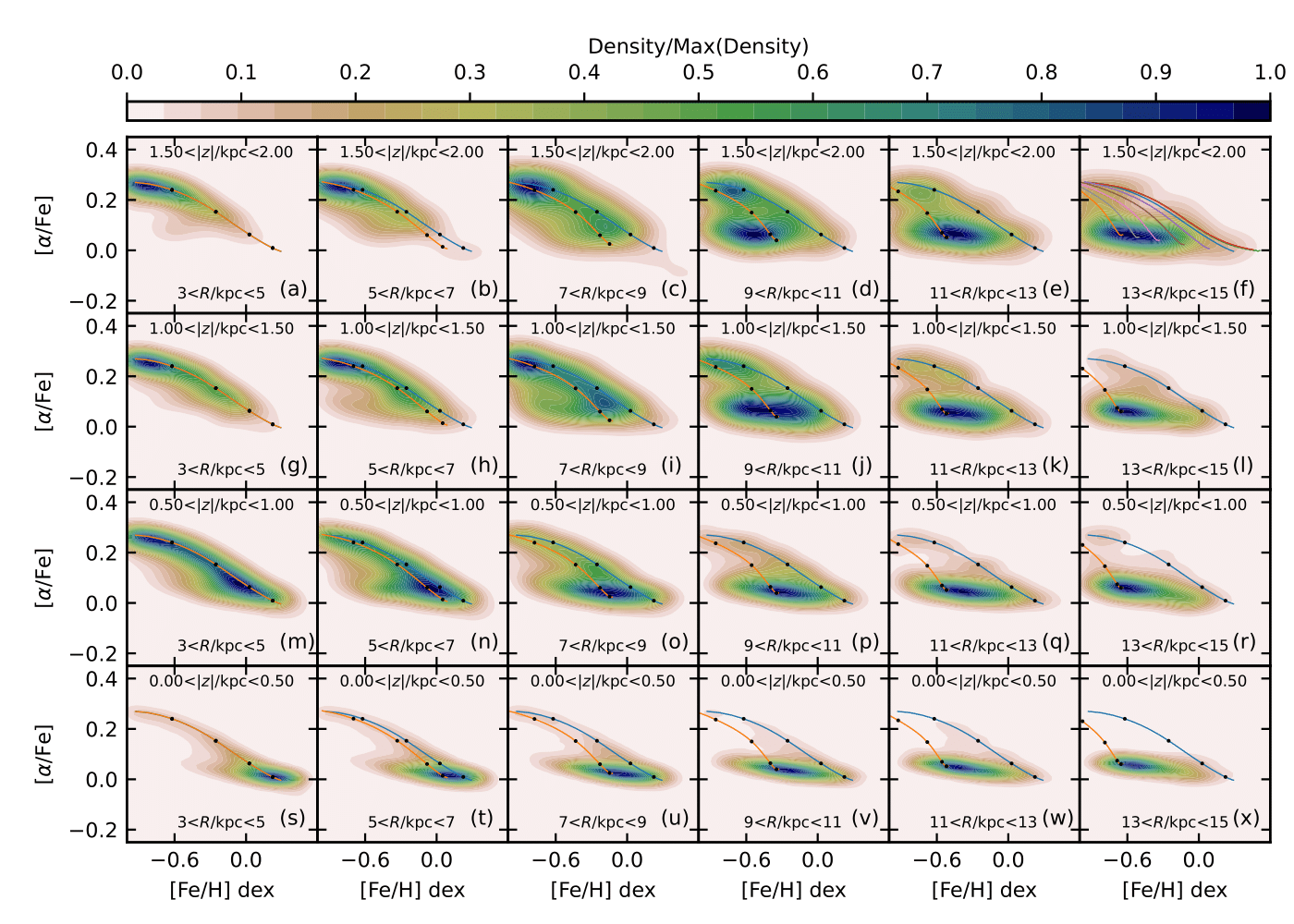}
  \vspace*{-2mm}
  \caption{The distribution of stars in the $([\alpha/{\rm Fe}], [{\rm Fe/H}])$ plane at different Galactic locations for our fiducial model as predicted by the selection function from \cite{2021MNRAS.507.5882S}. \alphafe{} ([Mg/Fe]) has been scaled so that the highest value is consistent with the observed value in Figure \ref{fig:rz_var_apogee}. The tracks from our fiducial model are shown in the same fashion as Figure \ref{fig:rz_var_apogee}. }
  \label{fig:rz_var_fiducial}
\end{figure*}

\begin{figure*}
\centering
  \includegraphics[width=0.95\linewidth]{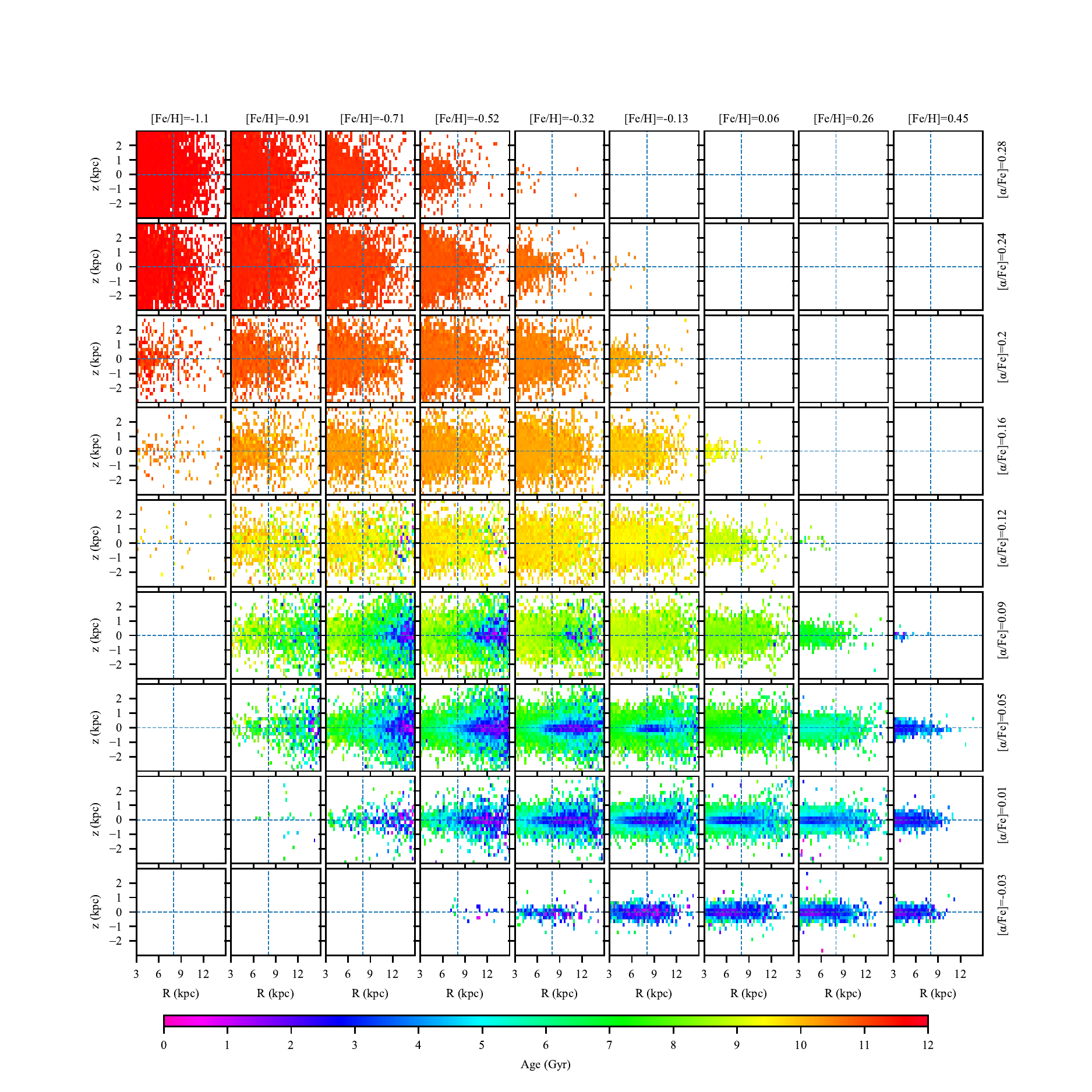}
  \vspace*{-2mm}
  \caption{The median age distribution in the ($R$, $z$) plane in our fiducial model for mono-abundance populations (MAPs). Each panel corresponds to a small region on the two-dimensional \fehalpha{} plane whose median \feh{} and \alphafe{} are listed on the top of each column and the right of each row. Within each panel (MAP), we show the median age at various ($R$, $z$) positions.}
  \label{fig:rz_map_age}
\end{figure*}

\begin{figure*}
\centering
  \includegraphics[width=0.95\linewidth]{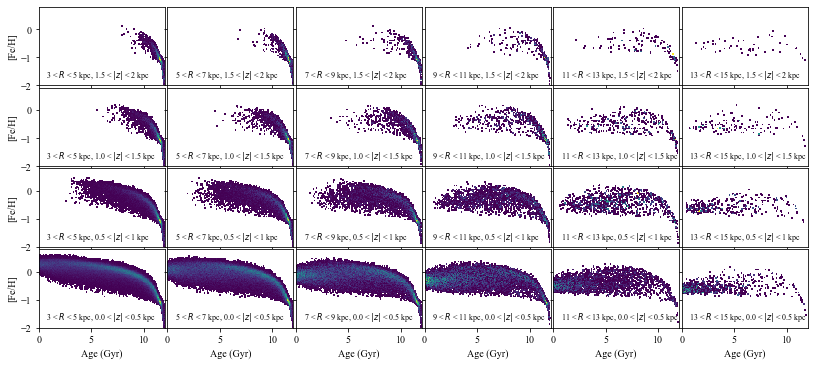}
  \vspace*{-2mm}
  \caption{Age-metallicity relation for different ($R$, $|z|$) positions in the fiducial model. Each column corresponds to a range in Galactocentric radius and each row corresponds to a range in the height from the Galactic plane.  }
  \label{fig:rz_feh_age}
\end{figure*}

\subsection{SNe Ia} \label{result: snia}
As SNe Ia produce the majority of iron in nucleosynthesis, it is unsurprising that SNe Ia prescriptions will have a massive effect on our models. We explore two factors of SNe Ia with two separate pairs of models, the fraction of SNe Ia in the model and DTD. If we compare the fiducial model against model SNIa+ and SNIa- in Figure \ref{fig:tracks}, \ref{fig:alpha_vs_time}, and \ref{fig:fe_vs_time}, a small change to the fraction of SNe Ia is capable of dramatically shifting \alphafe{} in the model. When the fraction of SNe Ia is increased to 6.5\%, the evolution of [Mg/Fe] in Figure \ref{fig:alpha_vs_time} consistently leads the trajectory predicted by the analytic model, reaching at least 0.1 dex lower than the analytic equilibrium value. When SNe Ia fraction is lowered to 5\%, the opposite happens as the evolution of [Mg/Fe] lags behind the analytic model. The same trend is observed for \feh{} in Figure \ref{fig:fe_vs_time}. Therefore, modifying the fraction of SNe Ia can alter the rate of abundance evolution in the model as well as the final equilibrium values. The changes to the abundance levels after they stabilize are reflected in Figure \ref{fig:hist_2d} and \ref{fig:alpha_hist} where the equilibrium \feh{} and [Mg/Fe] of the low-\alphafe{} sequence vary. 

Although $f_{\rm SNIa}$ affects the rate of chemical evolution, it does not necessarily affect the \alphafe{}-dichotomy in our models. This is because $f_{\rm SNIa}$ also determines the final equilibrium values for \feh{} and \alphafe{} in our models, which are constrained by observations. if we compare the fiducial model and model SNIa+ in Figure \ref{fig:hist_2d} and \ref{fig:alpha_hist}, we can see that one percentage point difference in $f_{\rm SNIa}$ results in a shift of about 0.05 dex in [Mg/Fe]. We did not pick a value of $f_{\rm SNIa}$ below 5\% for model SNIa- for it would make the low-\alphafe{} sequence too high in \alphafe{}. Other parameters that affect the rate of chemical evolution, such as ejection and star formation history, are much less constrained by the Milky Way, and thus can play a greater role in shaping the \alphafe{}-dichotomy. 

The effect of the DTD is similar to SFE. When the total amount of gas is fixed by our infall scheme, the SFE determines the rate at which gas is enriched by stellar nucleosynthesis. As long as the model runs long enough, models with different SFEs will reach the same equilibrium abundance levels, but sections along the chemical evolutionary tracks will be populated by varying amounts of stars. Similarly, DTD does not alter the total amount of iron produced at the end in the model but only the rate at which iron is released into the model. When we examine the five diagnostic plots, DTD+ exhibits the same behaviour as SF- and DTD- as SF+. Our fiducial model increases the fraction SNe Ia in the beginning so that the fall in [Mg/Fe] can match that from the analytic model. However, a small adjustment in the SFE or the DTD can achieve the same effect. 

\subsection{($R$, $|z|$) variation} \label{result: rz_variation}
As explained in Section \ref{subsec:phase_space}, we are able to explore the density variation of the low- and high-\alphafe{} sequences in galactocentric ($R$, $|z|$) by making assumptions about the potential of the Milky Way. Figure \ref{fig:rz_var_apogee} and \ref{fig:rz_var_fiducial} show the ($R$, $|z|$) density variation in the [Fe/H]-\alphafe{} plane for the APOGEE survey and our fiducial model respectively. Figure \ref{fig:rz_var_apogee} is an updated version of the original famous Figure 4 from \cite{2015ApJ...808..132H}. Each column corresponds to a specific range of galactocentric distance from 3 kpc to a maximum of 15 kpc. Each row corresponds to a different range of $|z|$-height and extends as high as 2 kpc from the disk. We found reasonable qualitative agreement between observation and our fiducial model at all radii and $|z|$ as far as 1 kpc from the disk. Additionally, we are able to reproduce the \alphafe{} and \feh{} gradient in the low-\alphafe{} sequence in Figure \ref{fig:rz_var_apogee} and \ref{fig:rz_var_fiducial}. As we move farther away from the galactic centre, \alphafe{} slightly increases and \feh{} lowers. However, when $|z|$ is higher than 1 kpc, we found that the low-\alphafe{} sequence is much more pronounced than observed. 

We show the median age distribution across the ($R$, $z$) for different mono-abundance populations (MAPs) in Figure \ref{fig:rz_map_age}. \cite{2022MNRAS.512.2890L} made a similar figure from the APOGEE data, albeit limited by the lack of data in some panels. We are able to replicate the general trends shown in their Figure 7. The high-\alphafe{} populations are dominated by old stars regardless of their ($R$, $z$) positions. The intermediate-\alphafe{} populations have a mix of young and old stars, but old stars tend to dominate in the metal-rich end. And we see mostly young stars among the lowest-\alphafe{} populations. \cite{2022MNRAS.512.4697L} made a similar figure from a cosmological simulation but the sampling rate was too low to reveal any detail ($R$, $z$) variation, except for maybe the intermediate-\alphafe{} region. However, our model is able to resolve the detailed ($R$, $z$) distribution that is missing in their results. In the rows with \alphafe{}$\approx$0.01, 0.05, 0.09, the young stars become less and less visible until they suddenly dominate the most metal-rich panels. This is caused by the \alphafe{} gradient in the fiducial model shown in Figure \ref{fig:tracks}. The equilibrium value of \alphafe{} decreases with increasing \feh{} until 5 kpc where the trend reverses. This reversal makes this particular region in \fehalpha{} transitory for evolutionary tracks and thus lacks young stars. {However, there is one significant mismatch between our figure and the figure from \cite{2022MNRAS.512.2890L}. Their low-\alphafe{} MAPs are relatively old ($\approx$ 4 Gyr), while our low-\alphafe{} MAPs are much younger. This could suggest that an additional mechanism is required in our model to suppress star formation in the last four Gyr.   }

\cite{2019MNRAS.489.1742F} found that the age-metallicity is not monotonic in the solar neighbourhood and beyond. Metallicity increases with time until two to four Gyr ago and reverses. The exact trend depends on the specific ($R$, $|z|$) location in the Milky Way. They suggested that this trend could be due to the radial migration of stars. The in situ and old stars from other radii form the expected sequence where metallicity increases with time. A large amount of relatively young metal-rich stars from the inner zones migrate outwards, forming the expected turning point in the age-metallicity relation. \cite{2022MNRAS.512.4697L} found in a cosmological simulation that gas accretion and radial migration could reproduce a similar curved age-metallicity relation in a galaxy similar to the Galaxy. We show the ($R$, $|z|$) variation of the age-metallicity relation in our fiducial model in comparison to Figure 3 from \cite{2019MNRAS.489.1742F}. Although we are able to reproduce the trend well for the GALAH MSTO stars in the solar neighbourhood (see Figure \ref{fig:fe_vs_time}), we do not see the curve among young stars seen in APOGEE. There are two possible reasons for this. The first is that the young stars in the inner disk in our fiducial model need to radially migrate more towards the outer disk, as suggested by \cite{2019MNRAS.489.1742F}. The second is that the curve is an artificial feature caused by the large dispersion in stellar age estimates.  

\subsection{Summary} \label{result: summary}
In this section, we explored the effect of several key ingredients in our models. We found that the radial inflow of gas determines the amount of metallicity gradient we can predict in our models and radial migration does not significantly affect the chemical evolutionary tracks but has a tremendous effect on the chemical distribution in the solar neighbourhood. SFE, the fraction of SNe Ia, SNe Ia DTD, and supernovae ejection can all influence the rate of chemical enrichment, but SFE and SNe Ia DTD do not alter the final equilibrium abundance levels in our models, as long as the models have sufficient time to reach there. We also found that an early peak in star formation is essential to observing a high-\alphafe{} sequence and that the gap in \alphafe{} is a joint effort of low SFR due to exhaustion of gas and fast chemical evolution due to SNe Ia kicking off. Lastly, we showed that our fiducial model can qualitatively reproduce the (R, z) density variations reported by \cite{2015ApJ...808..132H} as well as the radial gradient in \feh{} and \alphafe{}.


\section{Comparison to existing literature} \label{sec:discussion}

The findings from this work resonate with the original model from \cite{2009MNRAS.396..203S}. 
In both models, there is no separate formation phase for the thick disk or the thin disk. {The wide metallicity spread in the solar neighbourhood is the result of stars born at different radii migrating to the solar neighbourhood. We extended the rationale to explain the two \alphafe{}-peaks.} The two \alphafe{}-sequences observed in the solar neighbourhood are due to the rapid fall of \alphafe{} and infall rate with time. The \alphafe{}-dichotomy in the solar neighbourhood is caused by the radial mixing of high-\alphafe{} stars from the inner disk and in situ low-\alphafe{} populations. This superposition of stellar populations born at different radii in the Galaxy creates the illusion of two distinct episodes of star formation at the solar radius. Since the chemical evolution only spends around two Gyr in the high-\alphafe{}, it is possible for the inner zones to form a substantial high-\alphafe{} sequence due to the high gas density and thus proportionally higher star formation rates. These high-\alphafe{} stars then radially migrate to other radii in the next ten Gyr. In this sense, these high-\alphafe{} stars are very similar to stars in the bulge.  

\begin{figure*}
  \includegraphics[width=0.6\linewidth]{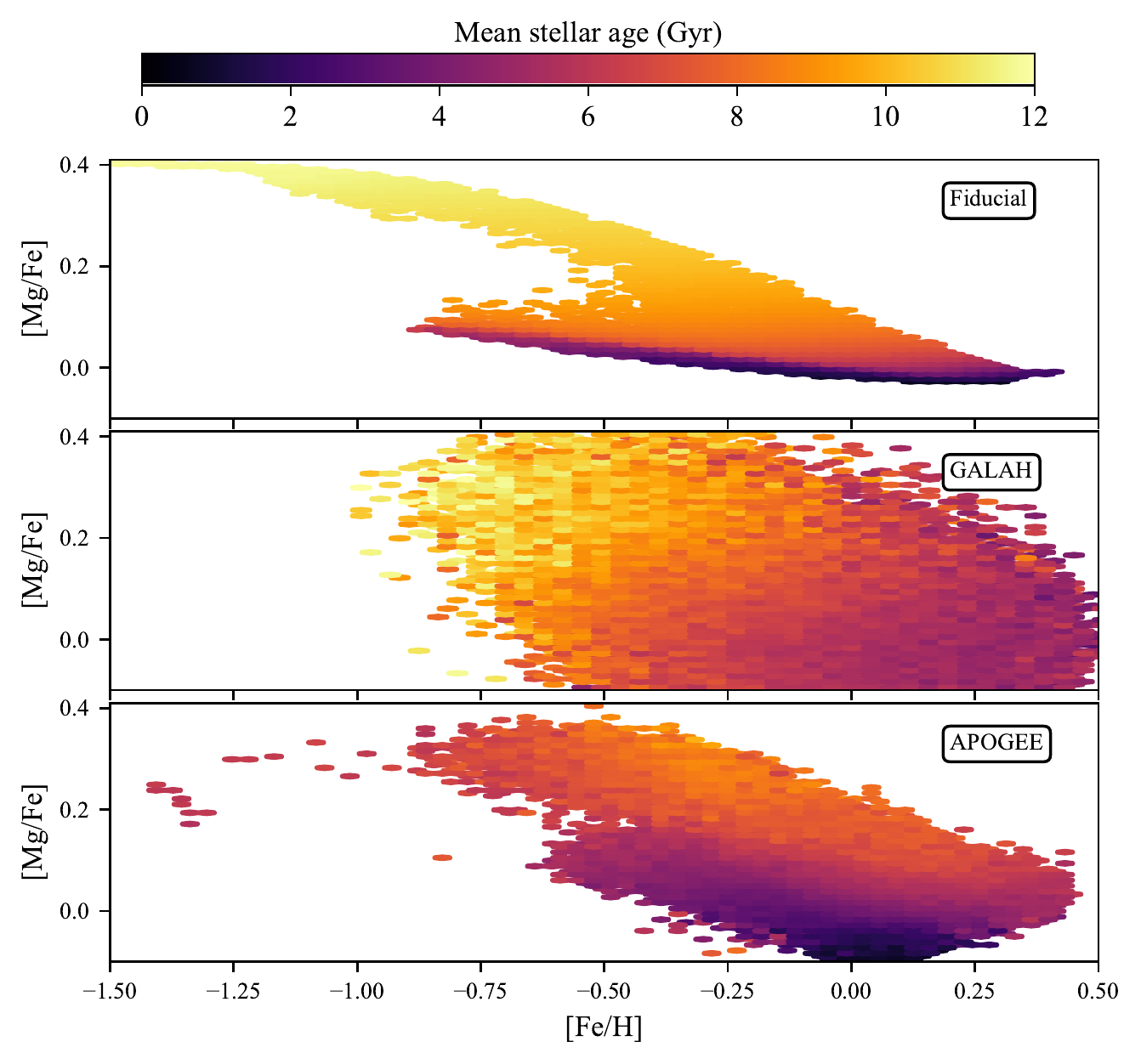}
  \centering
  \vspace*{-3mm}
  \caption{The mean stellar age in the (\feh{}, [Mg/Fe]) plane in the solar neighbourhood from our fiducial model (top) and the GALAH MSTO sample (middle), and APOGEE with AstroNN ages (bottom). }
  \label{fig:mean_age}
\end{figure*}

{Figure \ref{fig:mean_age} shows the distribution of mean stellar age in the (\feh{}, [Mg/Fe]) plane in the solar neighbourhood in our fiducial model, the GALAH MSTO sample, and the APOGEE sample with ages from the AstroNN value-added catalogue. The high-\alphafe{} stars are exclusively old in all panels, but their ages can significantly differ. Our fiducial model places the high-\alphafe{} stars at more than ten Gyr old and GALAH MSTO stars show similar old ages. AstroNN, however, measures high-\alphafe{} stars to be between six and eight Gyr old, much younger than our model and the GALAH MSTO sample. As for the low-\alphafe{} stars, there are few stars younger than four Gyr in the GALAH MSTO sample, while AstroNN confidently delivers a large number of young stars with ages as low as less than two Gyr. The features we see here in the data are most likely driven by the methodologies used to derive stellar ages which are still difficult to measure reliably. Nevertheless, the AStroNN ages reveal that the youngest stars that lie on the bottom of the low-\alphafe{} sequence along a ridge in the chemical space, similar to our fiducial model. However, this ridge does not extend to high metallicity in APOGEE, while it extends beyond \feh{}=0.25 in our model because we do not have a mechanism to suppress star formation in the inner zones in the last few Gyr. This same issue is observed in Figure \ref{fig:rz_map_age}. Because zones reach different equilibrium abundances, chemically distinct stars continue to form simultaneously in our model. These stars would then radially mix to drive the chemical scatter we observe at different locations in the Galaxy, including the solar neighbourhood. This is the key takeaway from this work. Even young stars with different metallicities or more generally different abundances at the same location do not necessarily have to form in-situ and thus a secondary infall episode resetting the chemistry in the two-infall model is not necessary. } 

In addition, we extend the comparison between data and our model beyond the solar neighbourhood. One of the arguments for a two-infall model or a quenching episode is that radial migration does not affect the inner zones of our Galaxy as much as the solar neighbourhood and thus does not explain the \alphafe{}-bimodality among the in situ stars in the bulge. In Figure \ref{fig:alpha_hist}, we showed that our SFH shaped by the infall history and a constant SFE is capable of reproducing the \alphafe{}-bimodality at 5 kpc. {We can also observe the \alphafe{}-bimodality between 3 and 5 kpc at intermediate-$|z|$ in our fiducial model in Figure \ref{fig:rz_var_fiducial}}. We found that the variation of stellar density in the \fehalpha{} plane as a function of Galactic ($R$, $|z|$) position predicted by our fiducial model agrees with APOGEE data in the solar neighbourhood and beyond in a direct comparison in Figure \ref{fig:rz_var_apogee} and \ref{fig:rz_var_fiducial}. Additionally, the enrichment histories of \feh{} and \alphafe{} as predicted by our fiducial model shown in Figure \ref{fig:alpha_vs_time} and \ref{fig:fe_vs_time} match the MSTO sample from GALAH DR3 with accurate ages and the analytic trajectories from \citet{2021MNRAS.507.5882S}.

\cite{2013A&A...558A...9M} incorporated the kinematics of particles from N-body simulations into a GCE model and their model is a pioneering hybrid model. Similar to \citet{2009MNRAS.396..203S}, they used a single star formation phase to replicate the properties of stars in the thin and thick disk. Their model is able to track the positions of particles more closely, while our model incorporates analytic prescriptions for the spatial distribution of stars over time. However, their model does not consider the radial flow of gas in the disk and the radial chemical gradient in their model is caused by the radially differential infall. Contrary to their model, we found that gas radial flow is the primary driver of a chemical gradient. Another key difference is that their model did not produce a disk with a two \alphafe{}-peaks and their \alphafe{} distribution is smooth. They argued that the \alphafe{}-dichotomy is caused by the survey selection effect. We now have significantly more data from large surveys and the selection effect cannot explain the \alphafe{}-dichotomy in the solar neighbourhood.

Our model is the closest to the model implemented in \cite{2015A&A...580A.126K}. Both of our models contain multi-phase ISM, radial gas inflow, radial migration of stars, and detailed nucleosynthesis. We both found that radial migration is needed to reproduce the chemical dispersion in the disk. However, similar to \cite{2013A&A...558A...9M}, their model did not reproduce a clear \alphafe{}-dichotomy or \alphafe{}-bimodality. Their work was before \cite{2015ApJ...808..132H} so the fit of their model on the \fehalpha{}-plane was considered sufficient at the time. It was difficult for them to constrain the parameters crucial to the evolution of \alphafe{} due to the lack of accurate stellar ages for a large sample.  We now know that \alphafe{} dropped rapidly over one to two Gyr rather than declining slowly over a long period of time (see the bottom panel of their Figure 10). This is one necessary condition for the \alphafe{}-dichotomy. 

The most recent two-infall model by \cite{2021A&A...647A..73S} compares their results to APOGEE DR16. They introduce a second infall about three to four Gyr after the initial infall forms the high-$\alpha$ sequence. The second infall episode dilutes the existing gas in the model and adjusts the abundance ratios to match the low-metallicity end in the low-$\alpha$ sequence. Although they were able to replicate the chemical spread in the disk with this approach, the evidence for such a dramatic chemical shift in the ISM remains elusive in the observed age-abundance relations in large-scale high-resolution spectroscopic surveys (see their Figure 14). Until then, models with smooth age-abundance relations should be preferred for studying the Milky Way. Our model not only reproduces the stellar density as a function of ($R$, $|z|$) positions in the \fehalpha{} plane (Figure \ref{fig:rz_var_fiducial}), but also matches the observations for the age-metallicity and age-\alphafe{} relationships (Figure \ref{fig:alpha_vs_time} and \ref{fig:fe_vs_time}). 

Our results resonate with some of those obtained from simulations. \cite{2019MNRAS.484.3476C} and \cite{2021MNRAS.501.5176K} both identified that the high- and low-\alphafe{} sequences should form under different SFRs. Our fiducial and alternative models showed that the rapid evolution of \alphafe{} caused by the onset of a large amount of SNe Ia as a result of intense star formation during the high-\alphafe{} regime and the rapid drop in the infall rate of fresh gas from the IGM are the key to replicating an \alphafe{}-dichotomy. However, the SFE, a fundamental parameter that governs star formation, does not necessarily need to vary to achieve the desired SFH, as long as it is sufficiently high. Even though the SFE can vary for a single galaxy in simulations, it is rare for it to change rapidly. The rest of the simulation works point to a large amount of gas with a different composition brought in through an accretion event. Although this scenario is feasible in theory, we are yet to identify the chemical signal of this event in the Milky Way. 

The SFH of our fiducial models is similar to those of \cite{2015A&A...578A..87S} and \cite{2018A&A...618A..78H} but the latter introduces quenching to create two distinct episodes of star formation. \cite{2018A&A...618A..78H} and we both found that a drop in the accretion rate during the formation of the low-\alphafe{} is necessary for \alphafe{}-bimodality in the inner zones. However, our SFE stays constant over time and our SFR does not drop to near zero and bounces back during the formation of the thin disk. Their model is a closed-box model and assumes that most gas accretion happened before any significant star formation. Their star formation rate does not depend on the amount of gas through Kennicutt-Schimidt law like ours and thus the SFE is their most important tool for manipulating the SFH. Our model, on the other hand, has continuous infall and outflow to shape the SFH and thus the role of the SFE is less crucial.

\cite{2020MNRAS.497.2371L} showed that a secondary infall and quenching are both necessary for an \alphafe{}-bimodality with their model. Our model does not explicitly model the bulge or the bar. However, we were able to replicate the \alphafe{}-bimodality between 3 and 5 kpc at $0.5<|z|<1$ kpc. Due to the radial exponential density of our gas disk, the inner zones of our fiducial model continue to form stars for the last five Gyr at a low SFR (our global SFR is 5-6 \msol per year in Figure \ref{fig:sfr}). If the \alphafe{}-bimodality is shown to be strong among stars close to the disk plane, we would have to modify our fiducial model to suppress star formation in the inner region. Nevertheless, instead of altering the SFE, a gas inflow mechanism similar to that implemented by \cite{2015A&A...580A.126K} can stop gas from flowing into the inner zones after the bar forms and induce a quenching in the SFR. 

\cite{2021MNRAS.508.4484J} also investigated the $(R,z)$ variation of \fehalpha{} distribution shown by \cite{2015ApJ...808..132H} in APOGEE. One of the important differences between our models is how they handle stellar evolution and nucleosynthesis. A detailed description of their model is presented in \cite{2020MNRAS.498.1364J}. They do not keep track of the lifetimes of stellar mass bins born at different time steps. Instead, they use a time-dependent parameter called cumulative return fraction to calculate how much mass from a single stellar population is returned to the ISM at a given time. This simplification can significantly affect the amount of new elements produced per solar mass. Our model instead uses the stellar remnant mass from the nucleosynthesis tables to keep track of the exact amount of each element.

The ($R$, $|z|$) variation in the \fehalpha{} distribution in their Figure 7 differs from APOGEE observations, even though they claimed to replicate the original result from \cite{2015ApJ...808..132H}. In Figure \ref{fig:rz_var_apogee} generated from APOGEE, we can see that the high-\alphafe{} sequence dominates the region with $|z| > 0.5$ kpc between 3 and 5 kpc. This behaviour is replicated by our model in Figure \ref{fig:rz_var_fiducial}, though our model predicts an excess of low-\alphafe{} stars at $0.5<|z|<1.0$ kpc. However, their Figure 7 shows a substantial amount of low-\alphafe{} stars in the inner zones regardless of $|z|$. The APOGEE data show that the high-\alphafe{} stars should be dominant at $|z| > 1$ kpc as far as 9 kpc from the Galactic centre, but their high-\alphafe{} sequence becomes far less visible beyond 5 kpc at any $|z|$. {Another issue is that they are unable to reproduce a clear \alphafe{}-gap.} In their Figure 12, they divided the stars by \feh{}, inflating the density of their high-\alphafe{} stars to present two \alphafe{}-peaks. One possible cause for the discrepancy is their high constant infall rate of 9 \msol{} per year which allows the formation of too many low-\alphafe{} stars late in the model. 

\section{Summary and future improvement} \label{sec:conclusion}
We have presented a multi-zone chemical evolution model of the Galaxy with the aim of understanding the general chemical abundance pattern of stars in the Milky Way. This model is an update of the important work by \citet{2009MNRAS.396..203S}, but makes use of a wealth of new observational constraints. Our model incorporates a number of important physical processes. It uses a two-component ISM consisting of a cold and a warm phase. Star formation is controlled by the infall rate of fresh gas and the Kennicut Schmidt law. The cold ISM participates in a radial flow, generating a chemical gradient in the disk. Stars are born out of cold gas, and upon death, they release newly synthesized elements into the warm ISM which gradually enriches the cold ISM over time. We are capable of tracking the number of major nucleosynthesis production sites from \cite{2020ApJ...900..179K}, including AGB, CCSN, and SNe Ia. Additionally, CCSN includes hypernovae (HNe), Type II supernovae (SNe II) and magneto-rotational supernovae (MRSNe). Finally, we model the present-day phase space distribution of stars in a dynamically consistent fashion using the prescription of \citet{2021MNRAS.507.5882S}. Our model is the first physical model derived from first principles (star formation and nucleosynthesis yields) to match many of the observational constraints for the chemical evolution of the Milky Way simultaneously-- the stellar distribution in \fehalpha{} plane at various $(R,z)$ locations, along with the age-\alphafe{} and age-\feh{} relations. Our main findings are given below.

\begin{itemize}

\item For the first time, we reproduced the variation of relative stellar density between the high-\alphafe{} and low-\alphafe{} sequences with Galactocentric radius $R$ and height from the disk plane $|z|$ that matches the APOGEE data. The intermediate-\alphafe{} gap can be clearly seen at locations where both \alphafe{}-sequences are visible.

\item Our model successfully reproduces the radial gradient of \feh{} and \alphafe{}. We show that the abundance gradient in the low-\alphafe{} sequence is primarily a consequence of the radial flow of enriched gas in the disc. A radially exponential gas disk with in situ infall is not sufficient to generate the observed abundance gradients.

\item Our model derived from the first principles is able to generate the same characteristic \alphafe{} evolution with time as \citet{2021MNRAS.507.5882S} and reinforce their results. Their analytic track is marked by a sharp fall of \alphafe{} with time due to the onset of SNe Ia supernovae and flat evolution afterwards. We identify three key properties that can alter the fall of \alphafe{} with time:  the fraction of stars that explode as SN Ia, the star formation rate early on, and finally lowering the time scale governing the delay time distribution of SN Ia. All three mechanisms have one thing in common- they increase the rate of SNe Ia per unit mass of the cold gas available for star formation. SNe Ia release primarily iron-peak elements. Increasing the rate of iron enrichment from this mechanism causes a faster the fall of \alphafe{}. 

\item In our model, the intermediate-\alphafe{} gap is caused by both the SFR and \alphafe{} rapidly declining. However, \citet{2021MNRAS.507.5882S} found that the rapid evolution of \alphafe{} alone can generate a gap in \alphafe{}, even with a constant star formation history. Our CSF model was not able to reproduce a prominent gap. One probable cause is that their analytic model took into account the inside-out growth of the disk, which could inflate the SFR in the inner disk in the early times. Unfortunately, our physical model is not yet stable enough to handle inside-out formation.

\end{itemize}

Our fiducial model has some differences with respect to observations. The first is that our high-\alphafe{} sequence is too metal-poor. At the beginning of our model, new elements are made exclusively by CCSN. Until SNe Ia take over as the primary source of iron, the only way to increase \feh{} is to increase the SFE or gas density in the model. We hold the SFE constant for the entire duration of our model so the only way is to increase the amount of gas in the first one Gyr. However, a huge amount of gas during this period would form an unrealistically large amount of metal-poor stars (\feh{} $< -1$). Another solution is to start our model with pre-enriched gas, but the chemical composition of this gas adds additional dimensions of freedom to our models.

Our replicated high-\alphafe{} sequence has a higher zero point of \alphafe{} at $\sim0.4$ compared to observations where the high-\alphafe{} sequence is at $\sim0.3$. A recent study of the halo stars suggests that there might be an early enrichment episode immediately after the Milky Way formed \citet{2022arXiv220402989C} that could explain this discrepancy. Another issue with our replicated high-\alphafe{} sequence is that its \feh{} range is not as extended as observed. Fragmentation of the early high-\alphafe{} disk \citep{2019MNRAS.484.3476C} and additional early accretion \citep{2022arXiv221101006C} could be the cause of the large \feh{} in the high-\alphafe{} sequence. The last detail we were not able to replicate at this point is the banana shape of the low-\alphafe{} sequence in the \fehalpha{}-plane. This feature of the Milky Way may be caused by the bar which is missing in our model.

There are still several areas for future improvement. One of them is to include inside-out growth of the disk, i.e., to allow for the scalelength of the disk to grow with time rather than being fixed. {For example, the radial distribution of gas infall in \cite{1997ApJ...477..765C} can change with time to achieve an inside-out growth of the gas disk.} Inside-out growth {concentrates early star formation in the inner region and thus more old and high-\alphafe{} stars will be found in the bulge region. The \alphafe{}-bimodality will be more pronounced in the bulge region. } Another area of improvement is related to better modelling of gas dynamical processes, e.g., the radial flow of gas, the infall of fresh gas, the multi-phase ISM, as well as star formation and its feedback on to the gas. Over the past few years, detailed cosmological hydrodynamical simulations of disc galaxies with star formation and feedback have made good progress in forming Milky-Way-like disc \citep{2020MNRAS.491.5435B,2018MNRAS.480..800H}; one could utilize these simulations to design more accurate prescriptions for gas dynamical processes. Currently, the results presented in this paper are restricted to [Fe/H] and \alphafe{}. In future, we plan to extend our model to include predictions for s- and r-process elements and {potentially derive their nucleosynthesis yields based on the observed age-abundance relation}. 

\section*{Acknowledgements}
B.C. is supported by the Research Training Program of the Australian Department of Education and the Hunstead student scholarship from the Dick Hunstead Fund. We thank Jianhui Lian for our extensive discussion of different GCE models in the literature. 

\bibliography{chemical_evolution}{}
\bibliographystyle{aasjournal}

\end{document}